\documentclass[aps,superscriptaddress,prl,floats,twocolumn,amsmath,amssymb,longbibliography,floatfix,preprintnumbers]{revtex4-2}

\usepackage{graphicx}
\usepackage{amsfonts}    
\usepackage{amssymb}     
\usepackage{amsbsy}      
\usepackage{amsmath}
\usepackage[colorlinks = true,
            linkcolor = magenta,
            urlcolor  = magenta,
            citecolor = magenta,
            anchorcolor = magenta]{hyperref}
\usepackage{xcolor}

\newcommand{\cdag}{c^\dagger}

\begin{document}

\title{Nontrivial entanglement passively mediated by a quenched magnetic impurity}

\author{A. A. Orlandini}
\affiliation {Instituto de F\'{\i}sica Rosario (CONICET) and Facultad de Ciencias Exactas, Ingeniería y Agrimensura (UNR)} 
\author{G. G. Blesio}
\affiliation {Instituto de F\'{\i}sica Rosario (CONICET) and Facultad de Ciencias Exactas, Ingeniería y Agrimensura (UNR)} 
\affiliation{Jo\v{z}ef  Stefan  Institute,  Jamova  39,  SI-1000  Ljubljana,  Slovenia}
\author{C. J. Gazza}
\affiliation {Instituto de F\'{\i}sica Rosario (CONICET) and Facultad de Ciencias Exactas, Ingeniería y Agrimensura (UNR)} 
\author{L. O. Manuel}
\affiliation {Instituto de F\'{\i}sica Rosario (CONICET) and Facultad de Ciencias Exactas, Ingeniería y Agrimensura (UNR)} 

\begin{abstract} 
We investigate the entanglement properties of a Kondo system undergoing a transition to a state with a quenched magnetic impurity, using the density 
matrix renormalization group (DMRG) method. We focus on a two-channel 
spin-1 Kondo impurity with single-ion anisotropy, where a quantum phase transition occurs between two topologically distinct local Fermi liquids. 
In the fully screened Kondo phase, realized at lower anisotropies, 
the entangled region surrounding the magnetic impurity mimics the Kondo 
screening cloud, although its length does not follow the conventional
behavior. In contrast, beyond the transition, the system enters a {\it non-Landau} Fermi liquid phase with a markedly different entanglement structure: 
as the impurity is quenched and disentangled from the rest of the system due to the breakdown of the Kondo effect, the two conduction channels \textemdash coupled only through the impurity\textemdash develop a significant degree of entanglement with one another. Our findings demonstrate that a quenched magnetic impurity can passively and efficiently mediate entanglement between spatially separated conduction bands. 
\end{abstract}

\maketitle 

{\textit{Introduction- }}  
The Kondo effect, a hallmark of condensed matter physics, manifests itself as an anomalous increase in the resistivity of metals containing magnetic impurities at low temperatures. The essence of the Kondo effect lies in the formation of a spin singlet between the impurity and the surrounding conduction electrons~\cite{Hewson1993}. 
This many-body phenomenon, whose influence extends far beyond its original context~\cite{Georges1996,Cronenwett1998,Madhavan1998,Kouwenhoven2001,Coleman2007,Gegenwart2008}, has revealed a rich landscape of physics, encompassing strong correlations, quantum criticality, and entanglement.  

Since its inception, the very existence and properties of the Kondo cloud —the electronic cloud that screens the spin of a magnetic impurity— have been the subject of intense debate~\cite{Sorensen1996,Barzykin1996,Affleck2001,Hand2006,Affleck2010, Yang2017,Mukherjee2022,Kim2024}. 
Recent experimental advancements have provided direct observations of the Kondo cloud, confirming its spatial extension over micrometer scales~\cite{Borzenets2020}. However, its internal structure remains poorly understood, prompting increasing interest in the distribution of entanglement within and beyond the cloud, as entanglement plays a crucial role in revealing its complex nature and making Kondo impurities ideal models for exploring the interplay between correlations and entanglement in many-body systems~\cite{Cho2006,Sorensen2007a,Bayat2012,Lee2015,Alkurtass2016,Alvarez2020,Kim2021,Shim2023,Nishikawa2025}.

Over time, more complex impurity systems involving multiple orbitals and conduction channels have emerged. Yet, the screening mechanisms and entanglement characteristics of these systems remain relatively unexplored. 
This growing complexity, enabled by advances in nanosystem development that allow for accurate modeling using multiorbital magnetic impurity models~\cite{Pustilnik2001,Potok2007,Parks2010,Minamitani2012,Evers2020,Blesio2024}, gives rise to diverse electronic states~\cite{Nozieres1980}.
For example,  topological quantum phase transitions (TQPTs) 
have been observed in certain impurity systems.  In nontrivial topological phases, entanglement measures—such as entanglement entropy  and concurrence—are essential for their characterization~\cite{Amico2008,Laflorencie2016}. 
These measures  also provide valuable insights into TQPTs, as they exhibit critical behavior in  specific 
entanglement quantities~\cite{WU-2004,Wagner2018}.

Recent studies have revealed the existence of a nontrivial topological Fermi
liquid in a spin-1 Kondo impurity coupled to two conduction channels~\cite{Blesio2018, Blesio2019,Zitko2021, Blesio2024}, for sufficiently large single-ion magnetic anisotropy $D$. This electronic phase is not adiabatically connected to the  non-interacting case, leading to its
characterization as a {\it non-Landau} Fermi liquid (NLFL). 
For negative $D$, fourth-order processes in the hybridization give rise to the usual Kondo effect~\cite{Blesio2018}.  While for positive $D$, it was found that a topological quantum phase transition occurs for a critical $D_c$ proportional to the Kondo temperature $T_K^0$, where the superscript refers to the case with $D=0$. This TQPT separates two different local Fermi liquids: for $D<D_c$ the spin of the impurity is screened by the usual Kondo effect~\cite{Blesio2019}, while for $D>D_c$ the magnetic degree of freedom of the impurity is quenched by the anisotropy.  Despite local moment phases of Kondo systems~\cite{Vojta2006,Wagner2018,Kim2024} have attracted considerable attention in the context of magnetic impurities in Dirac materials~\cite{Fritz2013} and superconductors~\cite{Moca2021}, quenched moment phases were previously considered trivial. 

To challenge this view, in this Letter we explore the spatial structure of entanglement of the spin-1 two-channel impurity model.  
Using DMRG and entanglement measures, we show how the multiorbital impurity, in the NLFL phase, serves as a passive mediator for the formation of significant entanglement between the two conduction baths, which interact indirectly through the impurity. This results in a nontrivial entanglement structure in the ground state. Additionally, we demonstrate that in the fully screened Kondo phase, the extent of the entanglement region is proportional to the size of the screening Kondo cloud, as it occurs in the $s=1/2$ single-channel Kondo model. However, these length scales do not follow the expected relationship, $\xi \propto 1/T_K$, indicating the presence of non-conventional Kondo physics.

\textit{Model and methods-} 
\begin{figure}[t!]
    \begin{center}
        \includegraphics*[width=1\columnwidth]{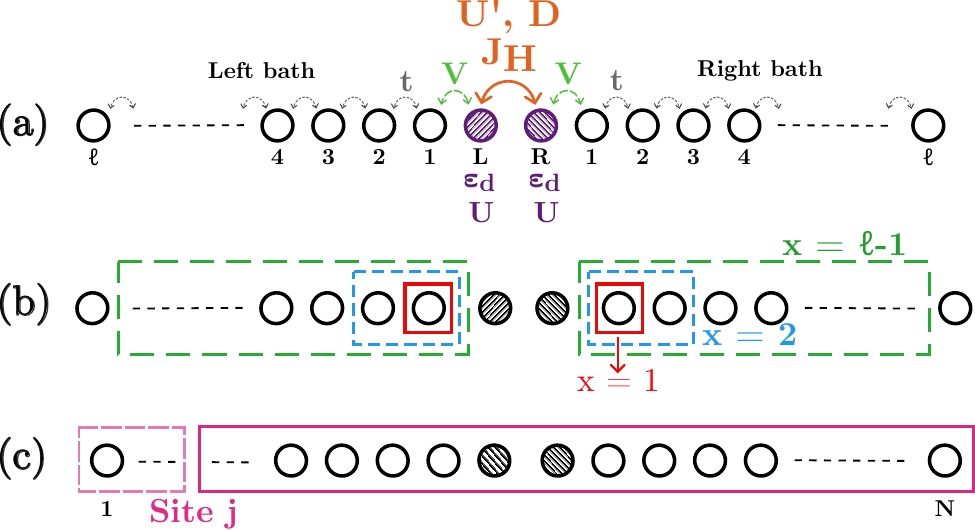}
        \caption{(a) Schematic representation of the 2CAM in a chain geometry. The two impurity orbitals, shown as central dashed circles, hybridize with the left and right conduction channels, defining the left and right orbitals, respectively.  
        (b) Illustration of the spatial coordinate $x$ used in  $\Sigma(x)$, measured symmetrically 
        with respect to the impurity. (c) System bipartition at 
        site $j$ for the von Neumann entropy computation.}
        \label{fig: modelo}
    \end{center}
\end{figure} 
We investigate the two-channel, two-orbital Anderson impurity model (2CAM) with degenerate impurity orbitals, each hosting, on average, one electron. A strong Hund coupling $J_H$
stabilizes a $S=1$ magnetic moment, while a positive single-ion anisotropy $D$ favors its quenching. Numerical simulations are performed in a chain geometry, as illustrated in Fig. \ref{fig: modelo}(a). The Hamiltonian is%
\begin{equation}
    \begin{split}
         H_{2CAM} &= H_{band} + H_{imp} + H_{hyb},
    \end{split}
    \label{ec: Hamiltoniano}
\end{equation}
where
\begin{equation}
    H_{band} = -t\sum_{i=1}^{\ell}\sum_{\alpha\sigma}  (\cdag_{i{\alpha}\sigma} c_{i+1\alpha\sigma} + {\rm H.c.}),
    \label{ec: HamiltonianoBand}
\end{equation}
\begin{equation}
    \begin{split}
        H_{imp} &= \sum_\alpha \varepsilon_d n_\alpha + U\sum_\alpha n_{\alpha\uparrow}n_{\alpha\downarrow} + U'n_{L} n_{R}    \\
        & - J_H\mathbf{S}_{L}\cdot\mathbf{S}_{R} +         D\left(S^z_L+S^z_R\right)^2,
    \end{split} 
    \label{ec: HamiltonianoImp}
\end{equation}
\begin{equation}
    H_{hyb} = V\sum_{\alpha\sigma}(\cdag_{1\alpha\sigma}d_{\alpha\sigma}+ {\rm H.c.}).
    \label{ec: HamiltonianoHyb}
\end{equation}
Here $d^\dag_{\alpha\sigma}$ ($\cdag_{i{\alpha}\sigma}$) creates an electron on the $\alpha=L,R$ impurity orbital (conduction band) with on-site impurity energy $\varepsilon_d$. 
The conduction channels in Eq.~\ref{ec: HamiltonianoBand} are tight-binding chains of length $\ell = \frac{N}{2}-1$, where $N$ is the total number of  ``sites'' (conduction channel and impurity orbitals).
We take the hybridization $V$  independent of the energy and the conduction channel.  In Eq. (\ref{ec: HamiltonianoImp}), $U$ is the intraorbital Coulomb repulsion energy, $U'$ corresponds to the interorbital repulsion, and $D$ is the single-ion anisotropy. 

To investigate the spatial dependence of the 2CAM, we computed the spin-spin correlations 
and analyzed the entanglement properties of our system using DMRG 
implemented through the 
ITensors library~\cite{ITENSOR-PAPER}.  DMRG is particularly well suited for studying the spatial properties of low-dimensional correlated models, as it naturally operates on the site basis of the Hilbert space. It has been extensively used in Kondo systems~\cite{Sorensen1996, Affleck2010, BUSSER-2010, RIBEIRO-2019, Hand2006, HOLZNER-2009, YANG-2005}. 
We performed calculations on systems with up to $N = 500$ sites, ensuring a maximum bond 
dimension sufficient to keep the truncation error below $10^{-9}$. 
The parameters were set to $\varepsilon_d = -U/2$ and $U' = J_H/4$, with $U = 1$. 
To enforce an impurity occupation forming a spin $S = 1$, 
we considered large values of the Hund coupling $J_H = 1.2 > W = 1$, where $W$ 
is the half-bandwidth of the conduction channels. 
The hybridization function  is $\Delta =  \pi V^2\rho_0(\varepsilon_F)$, where $\rho_0$ is the density of states of the conduction baths and 
we have taken $V \ll |\varepsilon_d|$ so that the system is in the Kondo regime with a good magnetic moment located in the impurity.

We complemented DMRG with Numerical Renormalization Group (NRG) calculations~\cite{Bulla2008}, specifically to detect abrupt changes in observables across the TQPT. The open-source NRG Ljubljana code~\cite{Zitko2009, nrglj} was used for these calculations. Technical details are provided in the Supplemental Material~\cite{SM}.
\nocite{SMHorodecki2009,SMWilson1975,SMRamsak2006}
\\

{\it Kondo screening cloud and entanglement entropy -}  

\begin{figure}[t!]
    \begin{center}
        \includegraphics*[width=1\columnwidth]{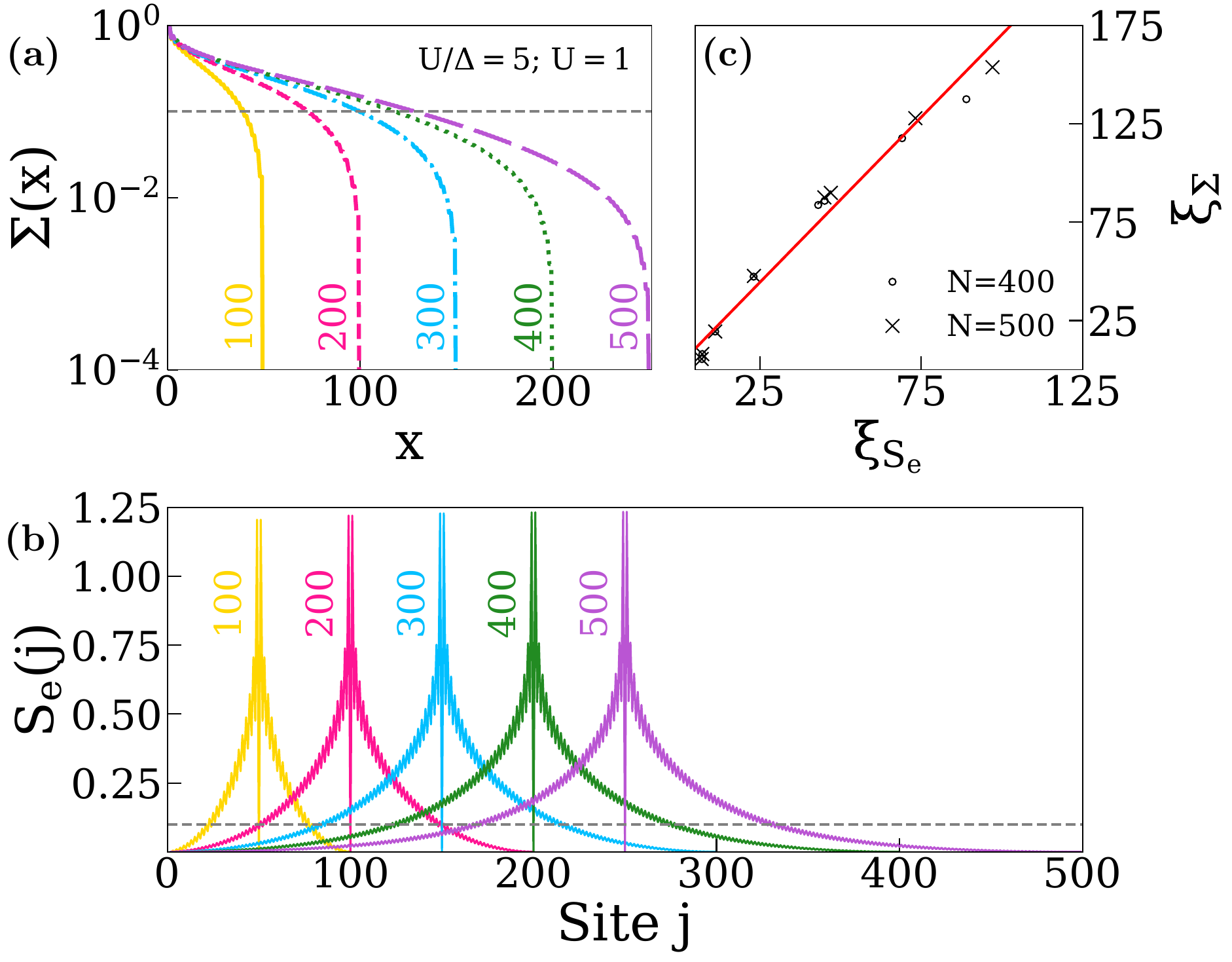}
        \caption{
        (a) Integrated spin-spin correlations $\Sigma$ for systems of different sizes. The threshold of $10\%$ used to extract $\xi_{\Sigma}(L)$ is indicated by the dashed horizontal line.  (b) Von Neumann's entropy taken for different bipartitions of the system for different system's size. 
        As in the upper panel, the threshold of the maximum value used to obtain $\xi_{S_e}(L)$ is shown. For both panels, 
        $\Delta = 0.2$ and $D=0$. (c) Kondo screening length $\xi_{\Sigma}$ versus $\xi_{S_e}$ for $N=400$ and $500$. The points are obtained for hybridizations $\Delta$ in [1/6; 5/4]. }
        \label{fig: longitud_CorrVsEnt}
    \end{center}
\end{figure}

One of the long-standing questions regarding the Kondo effect concerns the nature of the so-called Kondo cloud—the region surrounding the impurity where conduction electrons become correlated, forming a singlet with the impurity spin. A first estimate of its spatial extent is provided by the Kondo temperature $T_K$, as the Kondo screening length is expected to scale as $\xi_\Sigma \propto T_K^{-1}$~\cite{Affleck2010}. A more precise measure to probe the Kondo screening length is the integrated spin-correlation function, originally proposed in the context of the single-impurity Anderson model (SIAM)~\cite{HOLZNER-2009,Gubernatis1987,Costamagna2006} and extended to the 2CAM as:
\begin{equation}
    \Sigma(x)=1+\sum_{i=1}^x \frac{\langle \mathbf{S}_{L}\cdot\mathbf{s}_{i}\rangle}{\langle\mathbf{S}_{L}\cdot\mathbf{S}_{L}\rangle + \langle\mathbf{S}_{R}\cdot\mathbf{S}_{L}\rangle},
    \label{ec: sigma_function}
\end{equation}
where $x$ symmetrically spans the two chains coupled to the impurities [see schematic in Fig.~\ref{fig: modelo} {\color{blue}(b)}], and $\mathbf{S}_{\alpha}$ ($\mathbf{s}_{i}$) are the spin operators on the $\alpha=L,R$ impurity orbital (conduction band sites). Fig.~\ref{fig: longitud_CorrVsEnt} (a) shows the value of $\Sigma(x)$ for different chain lengths, at $D=0$. As more sites of the bath are taken into account (increased $x$), the integrated spin-correlation function goes towards zero, due to the singlet formation~\cite{SM}. 
Taking this result into account, the Kondo screening length is commonly 
defined as the distance at which the integrated spin-correlation function 
reaches 90$\%$ of its maximun value~\cite{HOLZNER-2009}, i.e., $\xi_\Sigma:\,\,\Sigma(\xi_\Sigma)=0.1\,\Sigma(x=0)$. 
As an alternative approach to characterizing the formation of the Kondo singlet, we can resort to entanglement properties. In particular, the von Neumann entanglement entropy of a bipartite system measures the entanglement between two complementary subsystems $A$ and $B$, and it is defined as $S(A)=-\sum_i \lambda_i \ln{\lambda_i}$, where $\lambda_i$ are the eigenvalues of the reduced density matrix $\rho$ of the subsystem $A$. 
By splitting the chain in the link between sites $j$ and $j+1$, defining the subsystems $A = [1,j]$ and $B=[j+1,N]$, we can study the entanglement $S(j)$ throughout the chain [see Fig.~\ref{fig: modelo}(c)]. 
We compute the impurity contribution to the entanglement entropy, defined as $S_e(j) \equiv S(j,V)-S(j,V=0)$, where $S(j,V=0)$ is the entanglement entropy with the impurity decoupled from the baths. 
In Fig.~\ref{fig: longitud_CorrVsEnt}(b), we display the entanglement entropy $S_e(j)$ for different chain length $N$, at $D=0$. 
These entanglement entropy plots are symmetric with 
respect to the center of the chain (i.e., the position of the impurity 
orbital), consistent with the left-right symmetry of the system. 
We define the entanglement length $\xi_{S_e}$ as the distance from the 
impurity at which the entanglement entropy drops to 10$\%$ of its maximum 
value. This length scale characterizes the extent of the region surrounding 
and including the impurity that is effectively disentangled from the rest of 
the system, thus providing an alternative measure of the Kondo singlet.

By comparing the screening length and the entanglement length across 
different hybridization strengths, we observe a linear relationship between 
the two, as shown in Fig~\ref{fig: longitud_CorrVsEnt}~(c). 
In the absence of anisotropy ($D = 0$), where the system exhibits a 
conventional Kondo effect, the spatial distribution of entanglement 
closely matches that of the Kondo cloud, regardless of the specific value 
of the Kondo temperature, as expected~\cite{Sorensen2007a}. This result establishes a direct connection 
between an experimentally accessible magnitude, as the Kondo screening cloud~\cite{Borzenets2020}, and an entanglement-based quantity, such as the von Neumann entanglement entropy~\cite{Lee2015,Shim2023}.

\begin{figure}[t!]
    \begin{center}
        \includegraphics*[width=1\columnwidth]{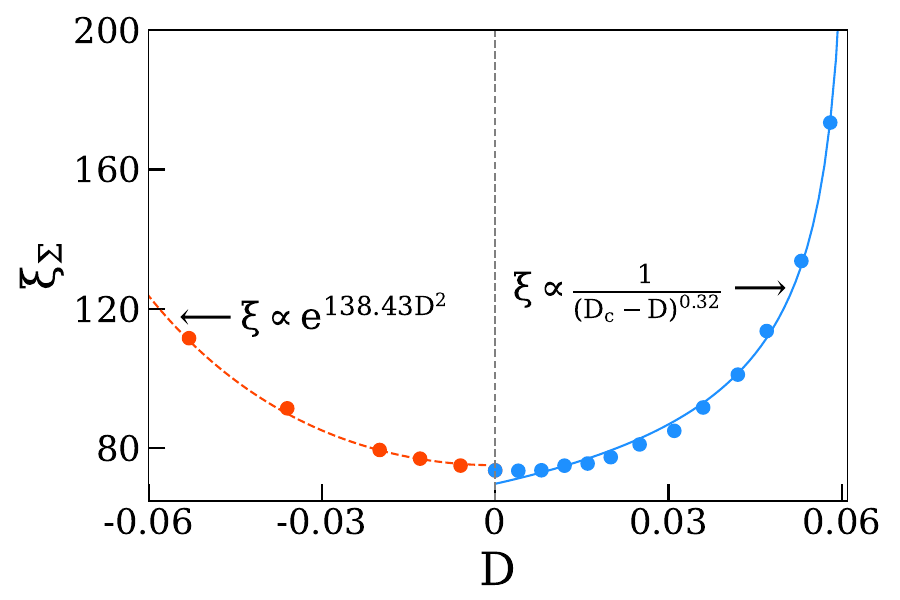}
        \caption{Screening length $\xi_{\Sigma}$ as a function of the single-ion anisotropy of the impurity. 
        Lines represent the relationship with the inverse of the Kondo temperature. At $D_c\sim 0.061$ occurs the TQPT. $\Delta = 1/4$ and $N = 400$.}
        \label{fig: longitud_anisotropia}
    \end{center}
\end{figure} 

When the anisotropy is introduced, distinct behaviors emerge depending on the values of $D$. Fig. \ref{fig: longitud_anisotropia} shows the screening length $\xi_\Sigma$ as a function of the single-ion anisotropy. 
For negative $D$, $\xi_\Sigma$ scales exponentially with $D^2$. 
Given that for negative $D$, $\ln T_K \propto -D^2$~\cite{Blesio2018}, in this case we recover the relation $\xi_\Sigma \propto 1/T_K$ for our multiorbital impurity model, consistent with results obtained for the fully screened $s=1/2$ Kondo phase~\cite{Sorensen1996,Affleck2010}. 
For positive $D$, it is know from Ref.~\cite{Zitko2021} that this model undergoes a TQPT at a finite critical anisotropy $D_c \simeq 2-3 T_K^{0}$, where
$T_K^0=T_K(D=0)$. For $D < D_c$, the Kondo temperature $T_K(D)$ follows an unusual power-like dependence $T_K(D) \propto T_K^{0}\left(\frac{D_c-D}{D_c}\right)^2.$ Consequently, we would expect that $\xi_{\Sigma} \propto (D_c-D)^{-2}$. However, we found that the screening length scales as 
\begin{equation}
    \xi_{\Sigma}\propto \left(\frac{D_c}{D_c-D}\right)^{0.32},
\end{equation}
diverging when $D \to D_c$ much more slowly than expected. The difference between the exponent of $\xi_\Sigma$ and $T_K^{-1}$ can be understood by using the relation $\xi_\Sigma = \frac{h\,v_F}{k_B\,T_K}$\cite{Affleck2010}. 
The slower increase of $\xi_\Sigma$ may indicate that the Fermi velocity $v_F$ is strongly renormalized by $D$, i.e., the band flattens as the system approaches the topological phase transition.  
As the anisotropy increases, the Kondo cloud begins to grow quickly with $D$, until the system is completely immersed in the cloud. On the other hand, within the {\it non-Landau} Fermi liquid phase, such a Kondo cloud does not exist at all as the impurity magnetic moment is quenched.  To further characterize this phase, we next analyze its entanglement structure. \\

\begin{figure}[t!]
    \begin{center}
        \includegraphics*[width=1\columnwidth]{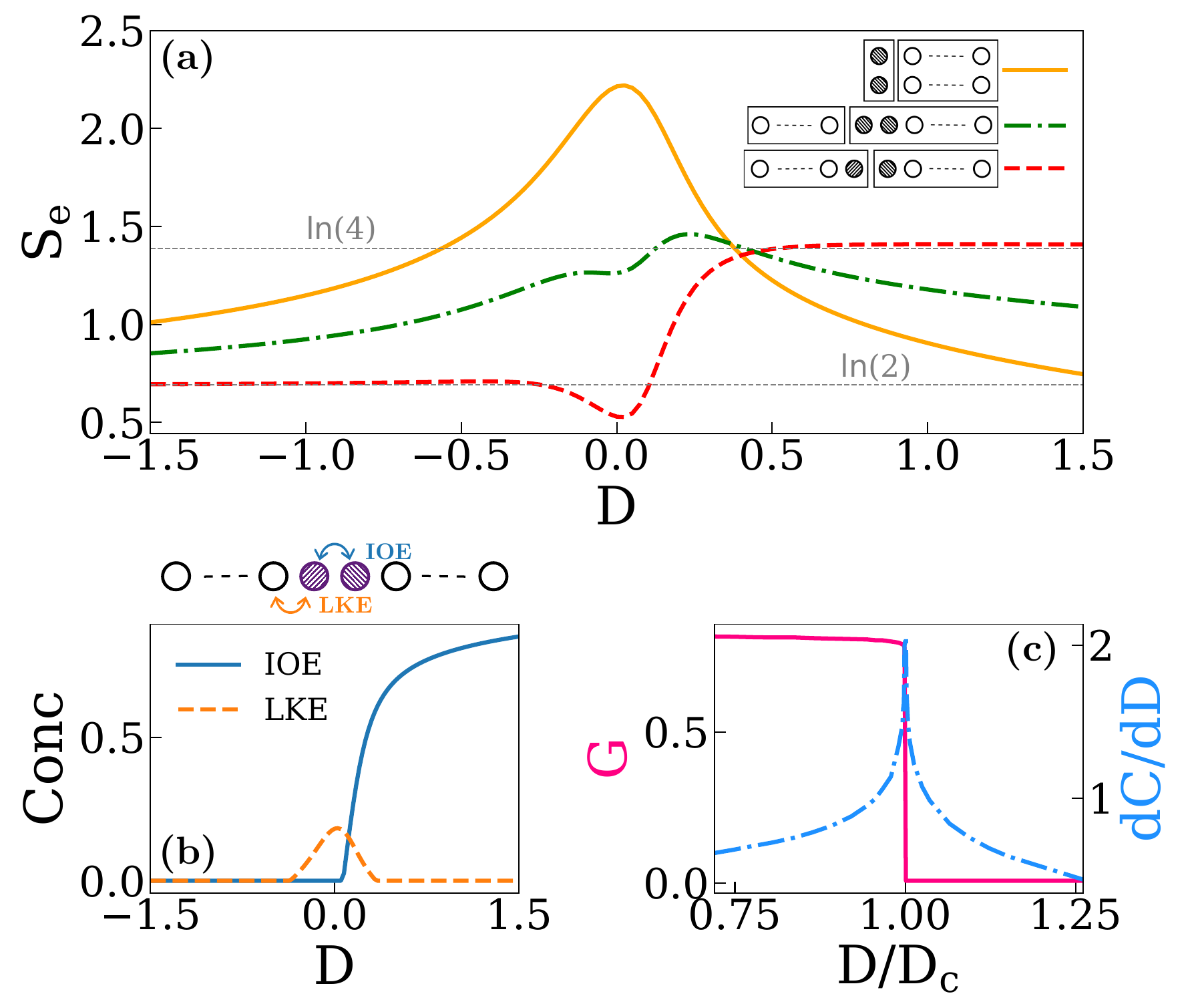}
        \caption{(a) Von Neumann entanglement entropy as a function of the impurity anisotropy for four different system partitions, as 
        indicated in the legend. $\Delta = 1/4$ and $N = 100$. (b) Concurrence for the inter orbital entanglement (IOE) and local Kondo entanglement (LKE), explained in the text. (c) Conductance $G$ and first derivative of the IOE concurrence at the TQPT critical point, computed using NRG at temperature $T = 10^{-14}$.}
        \label{fig: entanglement_measures}
    \end{center}
\end{figure}

{\it Entanglement structure-}  
In Fig.~\ref{fig: entanglement_measures} (a) we display the von Neumann's entanglement entropy $S_e$ against $D$ for a variety of bipartitions of the system. One of the most used bipartitions to study entanglement in condensed matter systems consists of separating the system in two identical parts. In our system, this consists of taking two blocks where each one has a conduction channel and its corresponding orbital of the impurity 
[see bottom bipartition sketch in Fig.~\ref{fig: entanglement_measures} (a)]. At large negative values of anisotropy $D$, $S_e$ tends to $\ln(2)$, indicating that both subsystems are entangled and that in each subsystem there is a degree of freedom that can take two possible values. This is in accordance with calculations using NRG where in this range the ground state of the system was found to be a Kondo singlet made up mainly of the impurity projections $M= \pm 1$ and corresponding states of the conduction baths~\cite{Blesio2019}. On the other hand,  $\ln (4)$ is obtained for larger positive $D$. 
To interpret this result, we employed a toy model (see Section B of the Supplemental Material~\cite{SM}) with a reduced number of degrees of 
freedom: each conduction bath and each impurity orbital were represented 
by a spin-$1/2$. For large positive anisotropies, this toy model has as 
its ground state a product state of two triplets with spin projection $S_z=0$: one formed by the two spins representing the impurity and the other by the two spins representing the conduction baths. Then, the ground state has the form
\begin{equation}
    |\psi_{D\gg 0}\rangle = \frac{1}{\sqrt{2}}(|\uparrow\downarrow\rangle+|\downarrow\uparrow\rangle)\otimes\frac{1}{\sqrt{2}}(|\uparrow\downarrow\rangle_c+|\downarrow\uparrow\rangle_c),
    \label{ec: GS_D_positivo}
\end{equation}
where the subscript \textit{c} notes the conduction baths. 
Since the entropy is calculated by cutting the system in half, two $\ln(2)$ arises from cutting both triplets. 
 Also, the toy model recovers the $\ln(2)$ limit of the entanglement entropy for $D\ll 0$ (see Fig. SM2 of Supplemental Material). 
Back to the 2CAM, for anisotropies well inside the non-Landau liquid regime where $S_e^{nL}=\ln(4)$, we expect that its ground state is composed of two triplets -one formed between both conduction baths and the other formed at the impurity- disentangled from each other.
Remarkably, the conduction channels are entangled although there are no direct interactions between them. Thus, the bath entanglement is passively mediated by the quenched impurity.  

When we take the impurity as one subsystem and the two conduction baths as the other [see top bipartition sketch in Fig.~\ref{fig: entanglement_measures} (a)], the aforementioned limit of $\ln(2)$ is recovered for $D\ll0$, while for $D\gg0$ the impurity and the conduction baths are disentangled, in agreement with the ground state of Eq.~(\ref{ec: GS_D_positivo}). Furthermore, we evaluated the entanglement entropy between a single conduction channel and the rest of the system [see the middle bipartition sketch in Fig.~\ref{fig: entanglement_measures} (a)]. In this case, we find that the entanglement entropy approaches $\ln(2)$ in both limits. For $D \ll 0$, this result supports the interpretation of the ground state forming a singlet by the two subsystems. For $D \gg 0$, the $\ln(2)$ contribution arises from the entanglement structure of the triplet state formed by both conduction channels. Overall, our findings demonstrate perfect agreement between the toy model and the 2CAM results in both asymptotic regimes, $D \ll 0$ and $D \gg 0$. 

To further deepen the understanding of the entanglement in our multiorbital Kondo impurity, we computed the concurrence~\cite{RAMSAK-06} between two electrons in different sites.
Specifically, the local Kondo entanglement (LKE), known as the entanglement for one of the impurity orbitals and his nearby corresponding conduction electron, and the inter orbital entanglement (IOE) between both of the impurity's orbitals~\cite{Li2017}. 
These calculations, shown in Fig. \ref{fig: entanglement_measures} (b), confirm the fact that in the {\it non-Landau} Fermi liquid phase the impurity orbitals are maximally entangled between them: IOE approaches the value of maximum concurrence, while LKE vanishes even before this occurs. These results are in accordance with those obtained for the toy model ground state (see Figure SM3 of Supplemental Material). 

Due to finite-size effects in the DMRG results, an abrupt phase transition 
cannot be resolved. Therefore, we employ the NRG method, which 
has been demonstrated to accurately identify the local TQPT~\cite{Blesio2018}.  
In Fig. \ref{fig: entanglement_measures} (c) we see the abrupt change in conductivity at the critical anisotropy $D_c$. This coincides with a change on the behavior of the IOE concurrence (or a peak in its first derivative). As expected, an entanglement measure successfully detects the topological transition. This means that, even though the concurrence is calculated exclusively between the two orbitals of the impurity, it still contains information about the full system. The transition can also be observed as the jump in the von Neumann entanglement entropy in the symmetric configuration [red dashed curve in Fig. \ref{fig: entanglement_measures} (a)], where it goes from low entanglement between the channels (due to large entanglement between the impurity and the bath), up to $ln(4)$ at high $D$ related to the two triplets as in Eq. \ref{ec: GS_D_positivo}.

\textit{Summary-} In this work, we show that the entangled region around a multiorbital magnetic impurity, in a fully Kondo screened phase, copies the Kondo screening cloud, whose length has recently been measured for the usual single-channel Kondo impurity~\cite{Borzenets2020}. We obtained this result by solving a $S=1$ Anderson impurity coupled to two conduction baths, by means of DMRG, which correctly captures the spatial properties of the system. In addition, when a single-ion anisotropy is turned on in the $S=1$ impurity, we have found that the ground state has 
 a nontrivial entanglement structure in its quenched impurity phase. 
 This structure consists in a product state where the impurity's orbitals are maximally entangled between them, forming a $S_z=0$ triplet, while simultaneously serving as a passive intermediary for the entanglement formation between the two conduction baths. This emergent entangled configuration 
 signals a deeper topological character of the {\it non-Landau} Fermi liquid phase. Our results thus uncover a novel entanglement signature of 
 topological Kondo breakdown, and suggest entanglement structure as a 
 powerful tool to probe nontrivial quantum phases beyond conventional 
 order parameters. \\

\textit{Acknowledgements-} 
A. A. O. thanks R. Picó for discussions and useful comments.
We acknowledge financial support by CONICET (Argentina) under Grant PIP No. 3220. Part of the results presented in this work have been obtained by using the facilities of the CCT-Rosario Computational Center, member of the High Performance Computing National System (SNCAD, MincyT- Argentina).
We also gratefully acknowledge the HPC RIVR consortium (\href{https://www.hpc-rivr.si}{www.hpc-rivr.si}) and EuroHPC JU (\href{https://eurohpc-ju.europa.eu/}{eurohpc-ju.europa.eu}) for funding this research by providing computing resources of the HPC system Vega at the Institute of Information Science (\href{https://www.izum.si/en/home/}{www.izum.si}).

\onecolumngrid
\newpage

\begin{center}

{\large\textbf{\boldmath
Supplemental Material\\ [0.5em] {\small for} \\ [0.5em]                                                                                                                                                                            
Nontrivial entanglement passively mediated by a quenched magnetic impurity}}\\[1.5em]

A. A. Orlandini, $^1$ G. G. Blesio,$^{1,2}$ C. J. Gazza,$^1$ and L. O. Manuel$^{1}$\\[0.5em]

\textit{\small
$^1$Instituto de F\'{\i}sica Rosario (CONICET) and Facultad de Ciencias Exactas, Ingenier\'{i}a y Agrimensura, 
Universidad Nacional de Rosario, 2000 Rosario, Argentina\\
$^2$Jo\v{z}ef Stefan Institute, Jamova 39, 1000 Ljubljana, Slovenia
}
\end{center}

\vspace{2em}

\begingroup
\centerline{\bfseries\large Abstract}
\smallskip
\begin{quotation}
\noindent
\textit{
We present the derivation of the $\Sigma(x)$ function, analyze the toy model 
used to interpret the DMRG results for the spin-1 two-channel Anderson model, 
and provide details of the numerical renormalization group (NRG) calculations.
}
\end{quotation}
\endgroup

\setcounter{figure}{0}
\renewcommand{\thefigure}{SM\arabic{figure}}

\renewcommand{\thesection}{\Alph{section}}%
\renewcommand{\thesubsection}{\Alph{section}.\arabic{subsection}}%
\renewcommand{\thesubsubsection}{ (\roman{subsubsection})}

\section*{A. $\Sigma(x)$ for 2CAM}
The total spin of the system is given by $\mathbf{S}_{tot}=\mathbf{S}_{imp}+\sum_{i}\mathbf{s}_i$, where $\mathbf{S}_{imp}=\mathbf{S}_L+\mathbf{S}_R$ is the spin of the two orbital impurity and $i$ runs over both chains (see Fig. 1(a) in the main text). Thus, the spin correlation between $\mathbf{S}_{tot}$ and the left impurity's orbital spin $\mathbf{S}_L$ can be decomposed as 
\begin{equation}
    \langle\mathbf{S}_{tot}\cdot\mathbf{S}_L\rangle=\langle\mathbf{S}_L\cdot\mathbf{S}_L\rangle+\langle\mathbf{S}_L\cdot\mathbf{S}_R\rangle+\langle\mathbf{S}_L\cdot\sum\mathbf{s}_i\rangle.
    \label{ec: SMproduct}
\end{equation}

In the Kondo phase with $D=0$ the ground state of the system is a singlet, so $\mathbf{S}_{tot}=0$ and the left-hand side of Eq. (\ref{ec: SMproduct}) vanishes. Reorganizing terms, it follows that
\begin{equation}
    1+\frac{\sum\langle\mathbf{S}_L\cdot\mathbf{s}_i\rangle}{\langle\mathbf{S}_L\cdot\mathbf{S}_L\rangle+\langle\mathbf{S}_L\cdot\mathbf{S}_R\rangle}=0,
\end{equation}
where the equality holds if the sum is performed taking all the sites of both conduction chains. If only the first $x$ sites of each chains are sum, one 
can define a function $\Sigma(x)$ that quantify the screening of the impurity
\begin{equation}
    \Sigma(x)=1+\sum_{i=1}^x \frac{\langle \mathbf{S}_{L}\cdot\mathbf{s}_{i}\rangle}{\langle\mathbf{S}_{L}\cdot\mathbf{S}_{L}\rangle + \langle\mathbf{S}_{L}\cdot\mathbf{S}_{R}\rangle}.
    \label{ec: SMsigma_function}
\end{equation}

Due to symmetry, Eq. (\ref{ec: SMsigma_function}) (Eq. (5) of the main text) can be equivalently defined with either $\mathbf{S}_L$ or $\mathbf{S}_R$ in the second term.

\section*{B. Toy model for the spin-1 Anderson model}

The spin-1 Anderson impurity model coupled to two conduction baths can be simplified by 
replacing each bath, as well as each orbital of the impurity, with a single spin-$1/2$ 
degree of freedom. The orbitals $a$ and $b$ of the impurity, replaced by the spin 
operators $\hat{\mathbf{S}}_a$ and $\hat{\mathbf{S}}_b$, respectively, are subject to a strong Hund's coupling $J_H$  yielding a total spin $S=1$. 
The impurity spin operators $\hat{\mathbf{S}}_a$ and $\hat{\mathbf{S}}_b$  are coupled via  exchange interactions $J_R$ and $J_L$ to spin-1/2 operators $\hat{\mathbf{s}}_R$ and $\hat{\mathbf{s}}_L$, representing the right and left conduction electrons, respectively. Additionally, the impurity has a single-ion anisotropy characterized by a parameter $D$. So the Hamiltonian of the resulting effective model takes the form:
\begin{equation}
 \hat{H} = J_L \hat{\mathbf{s}}_L \cdot \hat{\mathbf{S}}_a + J_R \hat{\mathbf{s}}_R \cdot \hat{\mathbf{S}}_b - J_H \hat{\mathbf{S}}_a \cdot \hat{\mathbf{S}}_b + D\left(\hat{S}^z_a+\hat{S}^z_b\right)^2,
\end{equation} 
where, as we said,  $\hat{\mathbf{S}}_a,\; \hat{\mathbf{S}}_b$ correspond to the impurity and $\hat{\mathbf{s}}_L, \hat{\mathbf {s}}_R$ to the left and right conduction channels, respectively. Since the total spin projection
\begin{equation}
 \hat{S}_{\rm tot}^z = \hat{s}^z_L + \hat{s}^z_R + \hat{S}^z_a + \hat{S}^z_b
\end{equation}
commutes with the Hamiltonian, it is possible to diagonalize it in each sector with a well-defined total spin projection. The ground state corresponds to the subspace with $S_{tot}^z=0$. For this subspace we define the base
\begin{equation}
    \begin{split}
        |1\rangle&\equiv|\uparrow\uparrow\downarrow\downarrow\rangle, \quad |2\rangle\equiv|\uparrow\downarrow\uparrow\downarrow\rangle, \quad |3\rangle\equiv|\downarrow\uparrow\downarrow\uparrow\rangle \\
        |4\rangle&\equiv|\downarrow\downarrow\uparrow\uparrow\rangle, \quad |5\rangle\equiv|\uparrow\downarrow\downarrow\uparrow\rangle, \quad |6\rangle\equiv|\downarrow\uparrow\uparrow\downarrow\rangle,
    \end{split}
\end{equation}
over which the Hamiltonian (for simplicity considering only the case $J_R = J_L \equiv J$) acts as
\begin{subequations}
    \begin{equation}
        \hat{H}|1\rangle  =  \frac{2 J+J_H}{4}|1\rangle - \frac{J_H}{2}|2\rangle,\\
    \end{equation}
    \begin{equation}
        \hat{H}|2\rangle  =  \frac{J_H-2 J}{4}|2\rangle -\frac{J_H}{2}|1\rangle + \frac{J}{2}\left(|5\rangle + |6\rangle\right), \\
    \end{equation}
    \begin{equation}
        \hat{H}|3\rangle  =  \frac{J_H-2 J}{4}|3\rangle -\frac{J_H}{2}|4\rangle +\frac{J}{2}\left(|5\rangle + |6\rangle\right),
    \end{equation}
    \begin{equation}
        \hat{H}|4\rangle  =  \frac{2 J+J_H}{4}|4\rangle - \frac{J_H}{2}|3\rangle,
    \end{equation}
    \begin{equation}
        \hat{H}|5\rangle  =  \left(D-\frac{2 J+J_H}{4}\right)|5\rangle + \frac{J}{2}\left(|2\rangle + |3\rangle\right),
    \end{equation}
    \begin{equation}
        \hat{H}|6\rangle  =  \left(D-\frac{2 J+J_H}{4}\right)|6 \rangle + \frac{J}{2}\left(|2\rangle+ |3\rangle\right).
    \end{equation}
\end{subequations}
Numerically, it is found that the ground state is generated by the states
\begin{equation}
    |a\rangle=\frac{|1\rangle+|4\rangle}{\sqrt{2}}, \quad |b\rangle=\frac{|2\rangle+|3\rangle}{\sqrt{2}}, \quad |c\rangle=\frac{|5\rangle+|6\rangle}{\sqrt{2}},
\end{equation}
i.e.,  
\begin{equation}
 |{\rm gs}\rangle = \alpha \left(\frac{ |\uparrow \uparrow \downarrow \downarrow\rangle +  |\downarrow\downarrow \uparrow\uparrow\rangle}{\sqrt{2}}\right) + \beta 
 \left(\frac{ |\uparrow\downarrow\uparrow \downarrow\rangle+|\downarrow \uparrow \downarrow \uparrow\rangle}{\sqrt{2}}\right) + \gamma \left(\frac{ |\uparrow \downarrow \downarrow \uparrow\rangle+
 |\downarrow \uparrow \uparrow \downarrow\rangle  }{\sqrt{2}}\right) 
 \label{coef}
\end{equation}

\begin{figure}
    \begin{center}
        \includegraphics[width=110mm]{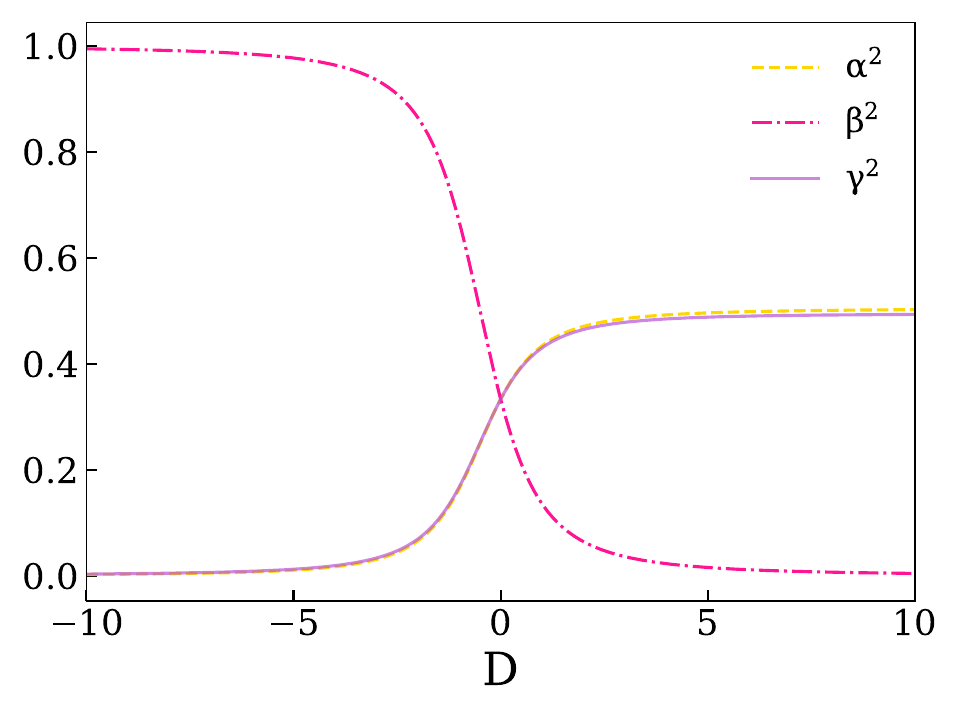}
        \caption{Coefficients $\alpha^2, \beta^2, \gamma^2$ (see equation \ref{coef}) of the wave function as a function of $D$. Parameters: $J=1, J_H=100$.}
        \label{wf_coeficientes}
    \end{center}
\end{figure}

Fig. \ref{wf_coeficientes} shows the dependence of the wave function coefficients on the anisotropy for $J=1, J_H=100$. It is found that for all anisotropy $\alpha \simeq \beta$. For these calculations, a very high value of the Hund interaction is taken so that the impurity is effectively a spin S=1.

If $D \ll -J$, $\gamma \simeq 1, \alpha \simeq \beta \simeq 0$ and the wave function consists mostly of states with projection $S^z=\pm 1$ of the impurity spins:
\begin{equation}
 |{\rm gs}(D \ll - J)\rangle = \frac{ |\uparrow \downarrow \downarrow \uparrow\rangle+
 |\downarrow \uparrow \uparrow \downarrow\rangle }{\sqrt{2}}.
 \label{gs4sneg}
\end{equation}

If $D \gg J$, then $\gamma \simeq 0,$ $\alpha \simeq \beta \simeq \frac{1}{\sqrt{2}}$. The projection $S^z=0$ of the two spins of the impurity is highly favored, resulting in a wave function
\begin{equation}
 |{\rm gs}\rangle = \frac{1}{2}\left(|\uparrow \uparrow \downarrow \downarrow\rangle + |\downarrow\downarrow \uparrow\uparrow\rangle +
 |\uparrow\downarrow\uparrow \downarrow\rangle+|\downarrow \uparrow \downarrow \uparrow\rangle\right).
 \label{gs4spos}
\end{equation}
This wave function is the product state of two triplets with zero spin projection, one formed between the two spins of the impurity and the other between the conduction spins $\mathbf{s_L}, \mathbf{s_R} $: 
\begin{equation} 
 |{\rm gs}(D\gg J)\rangle = |1,0\rangle_{\rm cond} \otimes |1,0\rangle_{\rm imp}.
 \label{gs4spos2}
\end{equation}

\subsection*{B.1. Entanglement entropy}
\subsubsection*{(i). A = left spin plus ``a'' spin of the impurity}

Let A be the block formed by the left conduction spin and the "a" spin of the impurity. Tracing out the rest of the system, we find that the reduced matrix of A is
\begin{equation}
    \begin{split}
 \hat{\rho}_A &= \frac{\alpha^2}{2} \left(|\uparrow \uparrow\rangle \langle \uparrow\uparrow| + |\downarrow\downarrow\rangle\langle \downarrow\downarrow|\right) + 
 \frac{\beta^2 + \gamma^2}{2}\left(|\uparrow\downarrow\rangle\langle \uparrow\downarrow| + |\downarrow\uparrow\rangle\langle\downarrow\uparrow|\right) \\
 &+ 
 \beta \gamma \left(|\uparrow\downarrow\rangle\langle\downarrow\uparrow| + |\downarrow\uparrow\rangle\langle\uparrow \downarrow|\right), 
    \end{split}
  \label{matA4s2}
 \end{equation}
where in the states $|\sigma\sigma'\rangle$, $\sigma$ corresponds to the left-hand spin and $\sigma'$ to the ``a'' spin of the impurity. This density matrix is expressed more compactly if we go from the states $|\sigma\sigma'\rangle$ to the triplets and singlets formed between both spins 1/2
\begin{equation}
 |\uparrow\uparrow\rangle = |1,1\rangle,\;\;\; |\downarrow\downarrow\rangle = |1,-1\rangle,\;\;\; |\uparrow\downarrow\rangle = \frac{|1,1\rangle + |0,0\rangle}{\sqrt{2}},\;\;\;
 |\downarrow\uparrow\rangle = \frac{|1,1\rangle-|0,0\rangle}{\sqrt{2}}.
\end{equation}
On this basis, the reduced diagonal matrix becomes diagonal
\begin{equation}
    \begin{split}
 \hat{\rho}_A &= \frac{\alpha^2}{2} \left( |1,1\rangle\langle 1,1| + |1,-1\rangle\langle 1,-1|\right) + \frac{(\beta + \gamma)^2}{2}|1,0\rangle \langle 1,0| \\
 &+ \frac{(\beta-\gamma)^2}{2}|0,0\rangle\langle 0,0|.
 \end{split}
 \label{matA4s}
\end{equation}
Finally, the entanglement entropy is~\cite{SMHorodecki2009}
\begin{equation}
 S_e= - \alpha^2 \ln \left[\frac{\alpha^2}{2}\right] - \frac{(\beta+\gamma)^2}{2}\ln\left[\frac{(\beta+\gamma)^2}{2}\right]
 -\frac{(\beta-\gamma)^2}{2}\ln\left[\frac{(\beta-\gamma)^2}{2}\right].
\end{equation}

\begin{figure}
    \begin{center}
        \includegraphics[width=110mm]{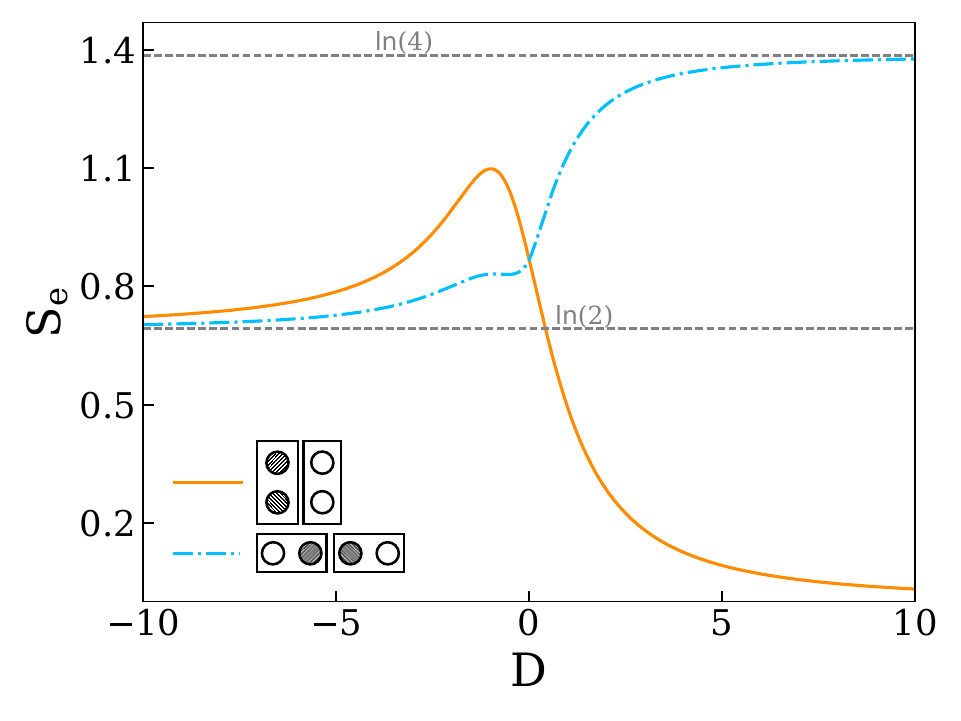}
        \caption{Entanglement entropy of the four spin model. Dot-dashed blue curve: block A contains the left spin and the ''a'' spin of the impurity. Solid orange curve: block A contains the two spins of the impurity. Same parameters as Fig. \ref{wf_coeficientes}.}
        \label{toyentropy_4s}
    \end{center}
\end{figure}

For this entropy. three particular cases can be analyzed:
\begin{itemize}
 \item If $D\gg J$, $\gamma \simeq 0, \alpha\simeq \beta \simeq 1/\sqrt{2}$. Therefore
 \begin{equation}
 S_e(D\gg J) = \ln 4.
\end{equation}
In this limit the reduced density matrix (\ref{matA4s}) is proportional to the identity: the four states of the Hilbert space of block A contribute equally, hence $\ln 4$. In a more physical and simple way, this can be interpreted by analyzing the shape of the ground state (\ref{gs4spos2}): by cutting the system in half, the two triplets are being cut, each of them contributing $\ln 2$ to the entropy.
 \item If $D=0$, $\alpha \simeq \beta \simeq 1/\sqrt{6}$, $\gamma \simeq \sqrt{2/3}$. Therefore, 
\begin{equation}
 S_e(D=0) \simeq \ln\left(\frac{4}{\sqrt{3}}\right).
\end{equation}
\item If $D \ll -J$ $\gamma \simeq 1, \alpha \simeq \beta \simeq 0$. The reduced density matrix (\ref{matA4s2}) has only equal contributions from the states $|\uparrow\downarrow\rangle, |\downarrow \uparrow\rangle$. Consequently the entropy is
\begin{equation}
 S_e (D\ll -J) = \ln 2.
\end{equation}
From the wave function (\ref{gs4sneg}) it follows that the states $|\uparrow\downarrow\rangle, |\downarrow\uparrow\rangle$ of both blocks, A and B, are entangled.
\end{itemize}
 
Fig.~\ref{toyentropy_4s} shows the entanglement entropy as a function of $D$. This behavior allows us to interpret the Anderson model results obtained using DMRG and presented in the main text.

\subsubsection*{(ii). A = spins ``a'' and ``b'' of the impurity}

Another partition of the system can be performed considering block A as the impurity spins and block B as the conduction spins. This partition is the most natural if we seek to understand how the
impurity is entangled with conduction baths in extended systems. The reduced density matrix is, in terms of the singlet and triplet states of block A (i.e., singlet and triplets of the impurity)
\begin{equation}
    \begin{split}
 \hat{\rho}_A &= \frac{\gamma^2}{2} \left( |1,1\rangle\langle 1,1| + |1,-1\rangle\langle 1,-1| \right) + \frac{(\beta + \alpha)^2}{2}|1.0\rangle \langle 1.0| \\
 &+ \frac{(\beta-\alpha)^2}{2}|0.0\rangle\langle 0.0|.
    \end{split}
 \label{mat4simp}
\end{equation}
Formally it has an expression similar to (\ref{matA4s2}): $\alpha \leftrightarrow \gamma$ are exchanged, however, the degrees of freedom of block A are different since they now correspond to the impurity spins.

The entanglement entropy is
\begin{equation}
 S_e= - \gamma^2 \ln \left[\frac{\gamma^2}{2}\right] - \frac{(\beta+\alpha)^2}{2}\ln\left[\frac{ (\beta+\alpha)^2}{2}\right]
 -\frac{(\beta-\alpha)^2}{2}\ln\left[\frac{(\beta-\alpha)^2}{2}\right].
\end{equation}
For $J_H \gg J$, $\alpha \simeq \beta,$ therefore, $2\alpha^2 = 1 -\gamma^2$ and the entropy reduces to
\begin{equation}
 S_e= - \gamma^2 \ln \left(\frac{\gamma^2}{2}\right) - (1-\gamma^2)\ln\left(1-\gamma^2\right)
\end{equation}
Repeating the analysis for limit values of $D$:
\begin{itemize}
 \item If $D \gg J$, $\gamma \simeq 0$, consequently
 \begin{equation}
  S_e (D\gg J) = 0,
 \end{equation}
which derives directly from the fact that the fundamental state (\ref{gs4spos2}) is a state product of an ''impurity'' state by a ``conduction'' state, there being no entanglement between both blocks.
\item If $D=0$, $\gamma^2 = \frac{2}{3}$ and
\begin{equation}
 S_e(D=0) = \ln 3.
\end{equation}
\item For $D=0$ and $J_H \gg J$ the ground state of the system is a spin 1 singlet between both blocks and its entropy then corresponds to $\ln 3$, for the three components that make up the singlet.
\item For $D \ll -J$, $\gamma \simeq 1$ and
\begin{equation}
 S_e(D\ll -J) = \ln 2.
\end{equation}
This result can be understood in a completely analogous way to what was done in the previous partition: the states $|\uparrow \downarrow\rangle, |\downarrow\uparrow\rangle$ of each block are intertwined.
\end{itemize}

The solid orange curve in Fig.~\ref{toyentropy_4s} shows $S_e$ as a function of $D$. This behavior is also obtained in the Anderson model in larger networks when using the ladder geometry.\\

\subsubsection*{(iii). A = left spin}

In this case the reduced matrix is proportional to the identity:
\begin{equation}
 \hat{\rho}_A = \frac{1}{2} \left(|\uparrow\rangle\langle\uparrow| + |\downarrow\rangle\langle\downarrow|\right).
\end{equation}

Consequently the entanglement entropy for all $D$ is
\begin{equation}
 S_e = \ln 2.
\end{equation}

\subsection*{B.2. Concurrence}

\begin{figure}
    \begin{center}
        \includegraphics[width=110mm]{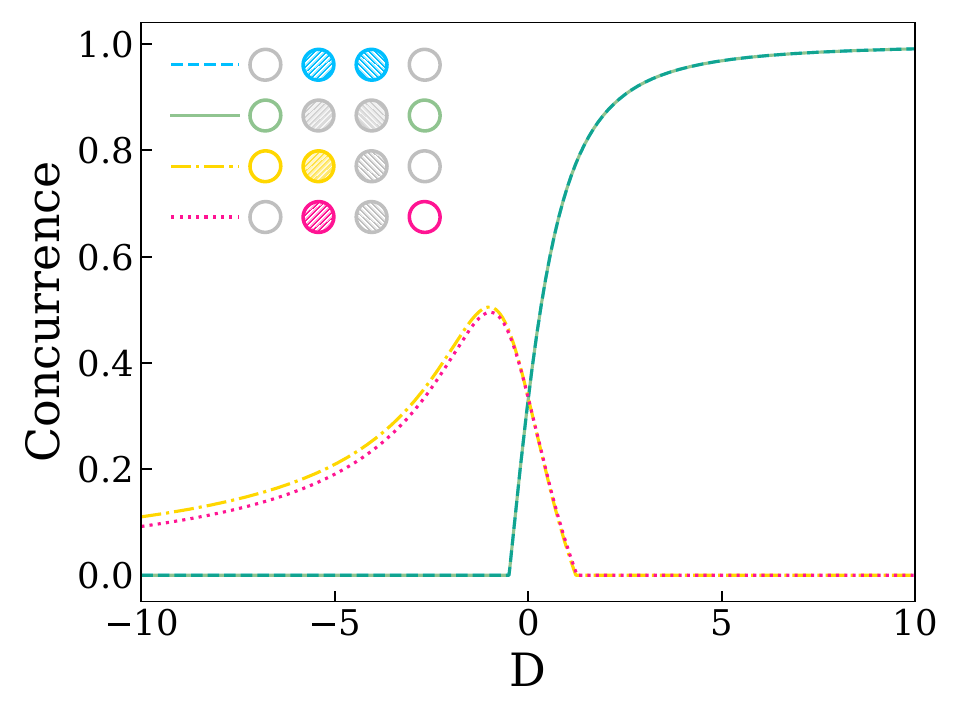}
        \caption{Concurrence for the four spin toy model. Spin pairs are color-coded in the legend; when two spins share a color, their corresponding concurrence curve is plotted in the same color. Same parameters as for Fig. \ref{wf_coeficientes}.}
        \label{toy_concurrence}
    \end{center}
\end{figure}

Concurrence is a measure of entanglement for systems composed of two subsystems of dimension 2 (qubits) and that works for both pure and mixed states. It is defined as~\cite{SMHorodecki2009}
\begin{equation}
    C(\rho_{AB})=\text{max}(0,\lambda_1-\lambda_2-\lambda_3-\lambda_4),
\end{equation}
where $\rho_{AB}$ is the reduced density matrix of the subsystem formed by qubits A and B, and the $\lambda_i^2$ are the eigenvalues, in decreasing order, of the operator
\begin{equation}
    R=\rho_{AB}(\sigma_y\otimes\sigma_y)\rho_{AB}^*(\sigma_y\otimes\sigma_y).
\end{equation}

Fig.~\ref{toy_concurrence} displays the concurrence for various spin-pair combinations. The dashed blue and green curves correspond to the entanglement between the impurity spins and those of the conduction bath, respectively. For sufficiently large anisotropy values, both curves approach the limit $C=1$, indicating maximal entanglement within these spin pairs. This behavior aligns with the interpretation that these pairs form triplet-like entangled states, which represent one of the maximally entangled configurations for two spins. Meanwhile, the cross-entanglement between impurity and bath spins—represented by the dot-dashed yellow and dotted red curves—vanishes entirely for large $D$, reinforcing the conclusion that the system’s ground state in the $D \gg J_H$ regime is a product state between the impurity and the bath subsystems.

\section*{C. Calculations on NRG}

The Numerical Renormalization Group~\cite{SMWilson1975,SMBulla2008} calculation where performed using the open source code ``NRG Ljubljana''~\cite{SMZitko2009,SMnrglj} for a Hamiltonian consisting of an impurity with two orbitals and two channels of conduction such as,
 
\begin{eqnarray} \label{ham}
H & = & \sum_{\alpha \sigma} \varepsilon_{d\alpha} d^\dagger_{\alpha\sigma} d_{\alpha\sigma} + 
U\sum_{\alpha}n_{d\alpha\uparrow}n_{d\alpha\downarrow} + U' n_{d_{a}}n_{d_{b}}-
\nonumber \\
 &  - & J_H {\vec S}_{d_{a}} \cdot {\vec S}_{d_{b}} + 
D S_z^2 +\label{and} \\
& + & \sum_{\nu k\alpha \sigma }\varepsilon _{k}c_{k\alpha \sigma }^{\dagger }c_{k\alpha \sigma }+ 
\sum_{k \alpha \sigma}\left( V_{\alpha} {c}^\dagger_{k \alpha\sigma}{d}_{\alpha \sigma} + {\rm H.c.}\right),
\nonumber 
\end{eqnarray}
where $\alpha$ represents the two orbitals $a$, $b$ for the impurity, or the two bands for the electronic bath, $\varepsilon_{d\alpha}$ is the on-site energy in the orbitals (same for both in our case), $U$ ($U'$) is the intra- (inter-) orbital Coulomb repulsion, $J_H$ the ferromagnetic Hund exchange, and $D$ the single-ion magnetic anisotropy. The hopping $V_{\alpha}$ characterizes the tunneling between the impurity orbitals and bath states at each channel $\alpha$, and we use (for both bath) a density of states for a chain with cosine dispersion, equivalent to the semicircular local DOS on the first site of the chain~\cite{SMnrglj}:
\begin{equation}
\rho(\omega)=\frac{2}{\pi W}\sqrt{1-\left(\frac{\omega}{W}\right)^2}\qquad for\qquad -W\leq\omega\leq W
\end{equation}  
where the half-bandwidth $W=1$.

For the calculations presented in Fig. 4 of the main text, we used $\varepsilon_d=-0.5$, $U=0.1$, $U'=0.3$, $J_H=-1.2$,$\Gamma=0.3562$ with $\Lambda=3$, up to 3000 NRG states and z-averaging with $N_z=4$. 

While not presented in this work, calculations based on a Kondo model—where the charge degree of freedom at the impurity is suppressed—yield similar results.

The concurrence between the two orbitals was calculated directly from spin-spin correlations and others expected values that are easily obtainable from NRG and was previously used for qubits~\cite{SMRamsak2006}.
\begin{equation}
\begin{split}
C=max(0,C_{\uparrow\downarrow},C_{||})/(P_{\uparrow\downarrow}+P_{||}),\quad~\, \\ 
C_{\uparrow\downarrow}=2\left|\langle S^+_a S^-_b \rangle\right|-2\sqrt{\langle P^\uparrow_a P^\uparrow_b \rangle\langle P^\downarrow_a P^\downarrow_b \rangle},\\
C_{\||}=2\left|\langle S^+_a S^+_b \rangle\right|-2\sqrt{\langle P^\uparrow_a P^\downarrow_b \rangle\langle P^\downarrow_a P^\uparrow_b \rangle}\\
\end{split}
\end{equation}
where $S^+_\alpha=\left(S^+_\alpha\right)^\dagger=d^\dagger_{\alpha\uparrow}d_{\alpha\downarrow}$, is the spin raising operators on the orbital $\alpha=a,b$, $P^\sigma_\alpha=n_{\alpha\sigma}(1-n_{\alpha\bar{\sigma}})$ is the operator projecting the state of orbital $\alpha$ into the space with one electron with spin $\sigma$. Finally, $P_{\uparrow\downarrow}=\langle P^\uparrow_a P^\downarrow_b + P^\downarrow_a P^\uparrow_b \rangle$ and $P_{||}=\langle P^\uparrow_a P^\uparrow_b + P^\downarrow_a P^\downarrow_b \rangle$ are the probabilities for antiparallel and parallel aligment of the spins at $a$ and $b$, respectively.

\bibliographystyle{apsrev4-2}

\begin{thebibliography}{58}%
\makeatletter
\providecommand \@ifxundefined [1]{%
 \@ifx{#1\undefined}
}%
\providecommand \@ifnum [1]{%
 \ifnum #1\expandafter \@firstoftwo
 \else \expandafter \@secondoftwo
 \fi
}%
\providecommand \@ifx [1]{%
 \ifx #1\expandafter \@firstoftwo
 \else \expandafter \@secondoftwo
 \fi
}%
\providecommand \natexlab [1]{#1}%
\providecommand \enquote  [1]{``#1''}%
\providecommand \bibnamefont  [1]{#1}%
\providecommand \bibfnamefont [1]{#1}%
\providecommand \citenamefont [1]{#1}%
\providecommand \href@noop [0]{\@secondoftwo}%
\providecommand \href [0]{\begingroup \@sanitize@url \@href}%
\providecommand \@href[1]{\@@startlink{#1}\@@href}%
\providecommand \@@href[1]{\endgroup#1\@@endlink}%
\providecommand \@sanitize@url [0]{\catcode `\\12\catcode `\$12\catcode
  `\&12\catcode `\#12\catcode `\^12\catcode `\_12\catcode `\%12\relax}%
\providecommand \@@startlink[1]{}%
\providecommand \@@endlink[0]{}%
\providecommand \url  [0]{\begingroup\@sanitize@url \@url }%
\providecommand \@url [1]{\endgroup\@href {#1}{\urlprefix }}%
\providecommand \urlprefix  [0]{URL }%
\providecommand \Eprint [0]{\href }%
\providecommand \doibase [0]{https://doi.org/}%
\providecommand \selectlanguage [0]{\@gobble}%
\providecommand \bibinfo  [0]{\@secondoftwo}%
\providecommand \bibfield  [0]{\@secondoftwo}%
\providecommand \translation [1]{[#1]}%
\providecommand \BibitemOpen [0]{}%
\providecommand \bibitemStop [0]{}%
\providecommand \bibitemNoStop [0]{.\EOS\space}%
\providecommand \EOS [0]{\spacefactor3000\relax}%
\providecommand \BibitemShut  [1]{\csname bibitem#1\endcsname}%
\let\auto@bib@innerbib\@empty
\bibitem [{\citenamefont {Hewson}(1993)}]{Hewson1993}%
  \BibitemOpen
  \bibfield  {author} {\bibinfo {author} {\bibfnamefont {A.~C.}\ \bibnamefont
  {Hewson}},\ }\href {https://doi.org/10.1017/CBO9780511470752} {\emph
  {\bibinfo {title} {The {K}ondo Problem to Heavy Fermions}}},\ Cambridge
  Studies in Magnetism\ (\bibinfo  {publisher} {Cambridge University Press},\
  \bibinfo {year} {1993})\BibitemShut {NoStop}%
\bibitem [{\citenamefont {Georges}\ \emph {et~al.}(1996)\citenamefont
  {Georges}, \citenamefont {Kotliar}, \citenamefont {Krauth},\ and\
  \citenamefont {Rozenberg}}]{Georges1996}%
  \BibitemOpen
  \bibfield  {author} {\bibinfo {author} {\bibfnamefont {A.}~\bibnamefont
  {Georges}}, \bibinfo {author} {\bibfnamefont {G.}~\bibnamefont {Kotliar}},
  \bibinfo {author} {\bibfnamefont {W.}~\bibnamefont {Krauth}},\ and\ \bibinfo
  {author} {\bibfnamefont {M.~J.}\ \bibnamefont {Rozenberg}},\ }\bibfield
  {title} {\bibinfo {title} {Dynamical mean-field theory of strongly correlated
  fermion systems and the limit of infinite dimensions},\ }\href
  {https://doi.org/10.1103/RevModPhys.68.13} {\bibfield  {journal} {\bibinfo
  {journal} {Rev. Mod. Phys.}\ }\textbf {\bibinfo {volume} {68}},\ \bibinfo
  {pages} {13} (\bibinfo {year} {1996})}\BibitemShut {NoStop}%
\bibitem [{\citenamefont {Cronenwett}\ \emph {et~al.}(1998)\citenamefont
  {Cronenwett}, \citenamefont {Oosterkamp},\ and\ \citenamefont
  {Kouwenhoven}}]{Cronenwett1998}%
  \BibitemOpen
  \bibfield  {author} {\bibinfo {author} {\bibfnamefont {S.~M.}\ \bibnamefont
  {Cronenwett}}, \bibinfo {author} {\bibfnamefont {T.~H.}\ \bibnamefont
  {Oosterkamp}},\ and\ \bibinfo {author} {\bibfnamefont {L.~P.}\ \bibnamefont
  {Kouwenhoven}},\ }\bibfield  {title} {\bibinfo {title} {{A Tunable Kondo
  Effect in Quantum Dots}},\ }\href
  {https://doi.org/10.1126/science.281.5376.540} {\bibfield  {journal}
  {\bibinfo  {journal} {Science}\ }\textbf {\bibinfo {volume} {281}},\ \bibinfo
  {pages} {540} (\bibinfo {year} {1998})}\BibitemShut {NoStop}%
\bibitem [{\citenamefont {Madhavan}\ \emph {et~al.}(1998)\citenamefont
  {Madhavan}, \citenamefont {Chen}, \citenamefont {Jamneala}, \citenamefont
  {Crommie},\ and\ \citenamefont {Wingreen}}]{Madhavan1998}%
  \BibitemOpen
  \bibfield  {author} {\bibinfo {author} {\bibfnamefont {V.}~\bibnamefont
  {Madhavan}}, \bibinfo {author} {\bibfnamefont {W.}~\bibnamefont {Chen}},
  \bibinfo {author} {\bibfnamefont {T.}~\bibnamefont {Jamneala}}, \bibinfo
  {author} {\bibfnamefont {M.~F.}\ \bibnamefont {Crommie}},\ and\ \bibinfo
  {author} {\bibfnamefont {N.~S.}\ \bibnamefont {Wingreen}},\ }\bibfield
  {title} {\bibinfo {title} {{Tunneling into a Single Magnetic Atom:
  Spectroscopic Evidence of the Kondo Resonance}},\ }\href
  {https://doi.org/10.1126/science.280.5363.567} {\bibfield  {journal}
  {\bibinfo  {journal} {Science}\ }\textbf {\bibinfo {volume} {280}},\ \bibinfo
  {pages} {567} (\bibinfo {year} {1998})}\BibitemShut {NoStop}%
\bibitem [{\citenamefont {Kouwenhoven}\ and\ \citenamefont
  {Glazman}(2001)}]{Kouwenhoven2001}%
  \BibitemOpen
  \bibfield  {author} {\bibinfo {author} {\bibfnamefont {L.}~\bibnamefont
  {Kouwenhoven}}\ and\ \bibinfo {author} {\bibfnamefont {L.}~\bibnamefont
  {Glazman}},\ }\bibfield  {title} {\bibinfo {title} {{Revival of the Kondo
  effect}},\ }\href {https://doi.org/10.1088/2058-7058/14/1/28} {\bibfield
  {journal} {\bibinfo  {journal} {Physics World}\ }\textbf {\bibinfo {volume}
  {14}},\ \bibinfo {pages} {33} (\bibinfo {year} {2001})}\BibitemShut {NoStop}%
\bibitem [{\citenamefont {Coleman}(2007)}]{Coleman2007}%
  \BibitemOpen
  \bibfield  {author} {\bibinfo {author} {\bibfnamefont {P.}~\bibnamefont
  {Coleman}},\ }\bibinfo {title} {{Heavy Fermions: Electrons at the Edge of
  Magnetism}},\ in\ \href
  {https://doi.org/https://doi.org/10.1002/9780470022184.hmm105} {\emph
  {\bibinfo {booktitle} {Handbook of Magnetism and Advanced Magnetic
  Materials}}}\ (\bibinfo  {publisher} {John Wiley \& Sons, Ltd},\ \bibinfo
  {year} {2007})\BibitemShut {NoStop}%
\bibitem [{\citenamefont {Gegenwart}\ \emph {et~al.}(2008)\citenamefont
  {Gegenwart}, \citenamefont {Si},\ and\ \citenamefont
  {Steglich}}]{Gegenwart2008}%
  \BibitemOpen
  \bibfield  {author} {\bibinfo {author} {\bibfnamefont {P.}~\bibnamefont
  {Gegenwart}}, \bibinfo {author} {\bibfnamefont {Q.}~\bibnamefont {Si}},\ and\
  \bibinfo {author} {\bibfnamefont {F.}~\bibnamefont {Steglich}},\ }\bibfield
  {title} {\bibinfo {title} {Quantum criticality in heavy-fermion metals},\
  }\href {https://doi.org/10.1038/nphys892} {\bibfield  {journal} {\bibinfo
  {journal} {Nature Physics}\ }\textbf {\bibinfo {volume} {4}},\ \bibinfo
  {pages} {186} (\bibinfo {year} {2008})}\BibitemShut {NoStop}%
\bibitem [{\citenamefont {S\o{}rensen}\ and\ \citenamefont
  {Affleck}(1996)}]{Sorensen1996}%
  \BibitemOpen
  \bibfield  {author} {\bibinfo {author} {\bibfnamefont {E.~S.}\ \bibnamefont
  {S\o{}rensen}}\ and\ \bibinfo {author} {\bibfnamefont {I.}~\bibnamefont
  {Affleck}},\ }\bibfield  {title} {\bibinfo {title} {{Scaling theory of the
  Kondo screening cloud}},\ }\href {https://doi.org/10.1103/PhysRevB.53.9153}
  {\bibfield  {journal} {\bibinfo  {journal} {Phys. Rev. B}\ }\textbf {\bibinfo
  {volume} {53}},\ \bibinfo {pages} {9153} (\bibinfo {year}
  {1996})}\BibitemShut {NoStop}%
\bibitem [{\citenamefont {Barzykin}\ and\ \citenamefont
  {Affleck}(1996)}]{Barzykin1996}%
  \BibitemOpen
  \bibfield  {author} {\bibinfo {author} {\bibfnamefont {V.}~\bibnamefont
  {Barzykin}}\ and\ \bibinfo {author} {\bibfnamefont {I.}~\bibnamefont
  {Affleck}},\ }\bibfield  {title} {\bibinfo {title} {{The Kondo Screening
  Cloud: What Can We Learn from Perturbation Theory?}},\ }\href
  {https://doi.org/10.1103/PhysRevLett.76.4959} {\bibfield  {journal} {\bibinfo
   {journal} {Phys. Rev. Lett.}\ }\textbf {\bibinfo {volume} {76}},\ \bibinfo
  {pages} {4959} (\bibinfo {year} {1996})}\BibitemShut {NoStop}%
\bibitem [{\citenamefont {Affleck}\ and\ \citenamefont
  {Simon}(2001)}]{Affleck2001}%
  \BibitemOpen
  \bibfield  {author} {\bibinfo {author} {\bibfnamefont {I.}~\bibnamefont
  {Affleck}}\ and\ \bibinfo {author} {\bibfnamefont {P.}~\bibnamefont
  {Simon}},\ }\bibfield  {title} {\bibinfo {title} {{Detecting the Kondo
  Screening Cloud Around a Quantum Dot}},\ }\href
  {https://doi.org/10.1103/PhysRevLett.86.2854} {\bibfield  {journal} {\bibinfo
   {journal} {Phys. Rev. Lett.}\ }\textbf {\bibinfo {volume} {86}},\ \bibinfo
  {pages} {2854} (\bibinfo {year} {2001})}\BibitemShut {NoStop}%
\bibitem [{\citenamefont {Hand}\ \emph {et~al.}(2006)\citenamefont {Hand},
  \citenamefont {Kroha},\ and\ \citenamefont {Monien}}]{Hand2006}%
  \BibitemOpen
  \bibfield  {author} {\bibinfo {author} {\bibfnamefont {T.}~\bibnamefont
  {Hand}}, \bibinfo {author} {\bibfnamefont {J.}~\bibnamefont {Kroha}},\ and\
  \bibinfo {author} {\bibfnamefont {H.}~\bibnamefont {Monien}},\ }\bibfield
  {title} {\bibinfo {title} {Spin correlations and finite-size effects in the
  one-dimensional {K}ondo box},\ }\href
  {https://doi.org/10.1103/PhysRevLett.97.136604} {\bibfield  {journal}
  {\bibinfo  {journal} {Phys. Rev. Lett.}\ }\textbf {\bibinfo {volume} {97}},\
  \bibinfo {pages} {136604} (\bibinfo {year} {2006})}\BibitemShut {NoStop}%
\bibitem [{\citenamefont {Affleck}(2010)}]{Affleck2010}%
  \BibitemOpen
  \bibfield  {author} {\bibinfo {author} {\bibfnamefont {I.}~\bibnamefont
  {Affleck}},\ }\bibinfo {title} {{The Kondo Screening Cloud: What It Is and
  How to Observe It}},\ in\ \href {https://doi.org/10.1142/9789814299442_0001}
  {\emph {\bibinfo {booktitle} {Perspectives of Mesoscopic Physics}}}\
  (\bibinfo  {publisher} {WORLD SCIENTIFIC},\ \bibinfo {year} {2010})\ pp.\
  \bibinfo {pages} {1--44}\BibitemShut {NoStop}%
\bibitem [{\citenamefont {Yang}\ and\ \citenamefont
  {Feiguin}(2017)}]{Yang2017}%
  \BibitemOpen
  \bibfield  {author} {\bibinfo {author} {\bibfnamefont {C.}~\bibnamefont
  {Yang}}\ and\ \bibinfo {author} {\bibfnamefont {A.~E.}\ \bibnamefont
  {Feiguin}},\ }\bibfield  {title} {\bibinfo {title} {Unveiling the internal
  entanglement structure of the kondo singlet},\ }\href
  {https://doi.org/10.1103/PhysRevB.95.115106} {\bibfield  {journal} {\bibinfo
  {journal} {Phys. Rev. B}\ }\textbf {\bibinfo {volume} {95}},\ \bibinfo
  {pages} {115106} (\bibinfo {year} {2017})}\BibitemShut {NoStop}%
\bibitem [{\citenamefont {Mukherjee}\ \emph {et~al.}(2022)\citenamefont
  {Mukherjee}, \citenamefont {Mukherjee}, \citenamefont {Vidhyadhiraja},
  \citenamefont {Taraphder},\ and\ \citenamefont {Lal}}]{Mukherjee2022}%
  \BibitemOpen
  \bibfield  {author} {\bibinfo {author} {\bibfnamefont {A.}~\bibnamefont
  {Mukherjee}}, \bibinfo {author} {\bibfnamefont {A.}~\bibnamefont
  {Mukherjee}}, \bibinfo {author} {\bibfnamefont {N.~S.}\ \bibnamefont
  {Vidhyadhiraja}}, \bibinfo {author} {\bibfnamefont {A.}~\bibnamefont
  {Taraphder}},\ and\ \bibinfo {author} {\bibfnamefont {S.}~\bibnamefont
  {Lal}},\ }\bibfield  {title} {\bibinfo {title} {{Unveiling the {Kondo} cloud:
  Unitary renormalization-group study of the Kondo model}},\ }\href
  {https://doi.org/10.1103/PhysRevB.105.085119} {\bibfield  {journal} {\bibinfo
   {journal} {Phys. Rev. B}\ }\textbf {\bibinfo {volume} {105}},\ \bibinfo
  {pages} {085119} (\bibinfo {year} {2022})}\BibitemShut {NoStop}%
\bibitem [{\citenamefont {Kim}\ \emph {et~al.}(2024)\citenamefont {Kim},
  \citenamefont {Shim}, \citenamefont {Sim},\ and\ \citenamefont
  {Kim}}]{Kim2024}%
  \BibitemOpen
  \bibfield  {author} {\bibinfo {author} {\bibfnamefont {M.~L.}\ \bibnamefont
  {Kim}}, \bibinfo {author} {\bibfnamefont {J.}~\bibnamefont {Shim}}, \bibinfo
  {author} {\bibfnamefont {H.~S.}\ \bibnamefont {Sim}},\ and\ \bibinfo {author}
  {\bibfnamefont {D.}~\bibnamefont {Kim}},\ }\href
  {https://arxiv.org/abs/2411.02723} {\bibinfo {title} {{Universal Spin
  Screening Clouds in Local Moment Phases}}} (\bibinfo {year} {2024}),\ \Eprint
  {https://arxiv.org/abs/2411.02723} {arXiv:2411.02723 [cond-mat.mes-hall]}
  \BibitemShut {NoStop}%
\bibitem [{\citenamefont {V.~Borzenets}\ \emph {et~al.}(2020)\citenamefont
  {V.~Borzenets}, \citenamefont {Shim}, \citenamefont {Chen}, \citenamefont
  {Ludwig}, \citenamefont {Wieck}, \citenamefont {Tarucha}, \citenamefont
  {Sim},\ and\ \citenamefont {Yamamoto}}]{Borzenets2020}%
  \BibitemOpen
  \bibfield  {author} {\bibinfo {author} {\bibfnamefont {I.}~\bibnamefont
  {V.~Borzenets}}, \bibinfo {author} {\bibfnamefont {J.}~\bibnamefont {Shim}},
  \bibinfo {author} {\bibfnamefont {J.~C.~H.}\ \bibnamefont {Chen}}, \bibinfo
  {author} {\bibfnamefont {A.}~\bibnamefont {Ludwig}}, \bibinfo {author}
  {\bibfnamefont {A.~D.}\ \bibnamefont {Wieck}}, \bibinfo {author}
  {\bibfnamefont {S.}~\bibnamefont {Tarucha}}, \bibinfo {author} {\bibfnamefont
  {H.-S.}\ \bibnamefont {Sim}},\ and\ \bibinfo {author} {\bibfnamefont
  {M.}~\bibnamefont {Yamamoto}},\ }\bibfield  {title} {\bibinfo {title}
  {Observation of the {Kondo} screening cloud},\ }\href
  {https://doi.org/10.1038/s41586-020-2058-6} {\bibfield  {journal} {\bibinfo
  {journal} {Nature}\ }\textbf {\bibinfo {volume} {579}},\ \bibinfo {pages}
  {210} (\bibinfo {year} {2020})}\BibitemShut {NoStop}%
\bibitem [{\citenamefont {Cho}\ and\ \citenamefont {McKenzie}(2006)}]{Cho2006}%
  \BibitemOpen
  \bibfield  {author} {\bibinfo {author} {\bibfnamefont {S.~Y.}\ \bibnamefont
  {Cho}}\ and\ \bibinfo {author} {\bibfnamefont {R.~H.}\ \bibnamefont
  {McKenzie}},\ }\bibfield  {title} {\bibinfo {title} {Quantum entanglement in
  the two-impurity {Kondo} model},\ }\href
  {https://doi.org/10.1103/PhysRevA.73.012109} {\bibfield  {journal} {\bibinfo
  {journal} {Phys. Rev. A}\ }\textbf {\bibinfo {volume} {73}},\ \bibinfo
  {pages} {012109} (\bibinfo {year} {2006})},\ \bibinfo {note}
  {http://arxiv.org/pdf/quant-ph/0509207}\BibitemShut {NoStop}%
\bibitem [{\citenamefont {Sørensen}\ \emph {et~al.}(2007)\citenamefont
  {Sørensen}, \citenamefont {Chang}, \citenamefont {Laflorencie},\ and\
  \citenamefont {Affleck}}]{Sorensen2007a}%
  \BibitemOpen
  \bibfield  {author} {\bibinfo {author} {\bibfnamefont {E.~S.}\ \bibnamefont
  {Sørensen}}, \bibinfo {author} {\bibfnamefont {M.-S.}\ \bibnamefont
  {Chang}}, \bibinfo {author} {\bibfnamefont {N.}~\bibnamefont {Laflorencie}},\
  and\ \bibinfo {author} {\bibfnamefont {I.}~\bibnamefont {Affleck}},\
  }\bibfield  {title} {\bibinfo {title} {{Impurity entanglement entropy and the
  Kondo screening cloud}},\ }\href
  {https://doi.org/10.1088/1742-5468/2007/01/L01001} {\bibfield  {journal}
  {\bibinfo  {journal} {Journal of Statistical Mechanics: Theory and
  Experiment}\ }\textbf {\bibinfo {volume} {2007}},\ \bibinfo {pages} {L01001}
  (\bibinfo {year} {2007})}\BibitemShut {NoStop}%
\bibitem [{\citenamefont {Bayat}\ \emph {et~al.}(2012)\citenamefont {Bayat},
  \citenamefont {Bose}, \citenamefont {Sodano},\ and\ \citenamefont
  {Johannesson}}]{Bayat2012}%
  \BibitemOpen
  \bibfield  {author} {\bibinfo {author} {\bibfnamefont {A.}~\bibnamefont
  {Bayat}}, \bibinfo {author} {\bibfnamefont {S.}~\bibnamefont {Bose}},
  \bibinfo {author} {\bibfnamefont {P.}~\bibnamefont {Sodano}},\ and\ \bibinfo
  {author} {\bibfnamefont {H.}~\bibnamefont {Johannesson}},\ }\bibfield
  {title} {\bibinfo {title} {{Entanglement Probe of Two-Impurity Kondo Physics
  in a Spin Chain}},\ }\href {https://doi.org/10.1103/PhysRevLett.109.066403}
  {\bibfield  {journal} {\bibinfo  {journal} {Phys. Rev. Lett.}\ }\textbf
  {\bibinfo {volume} {109}},\ \bibinfo {pages} {066403} (\bibinfo {year}
  {2012})}\BibitemShut {NoStop}%
\bibitem [{\citenamefont {Lee}\ \emph {et~al.}(2015)\citenamefont {Lee},
  \citenamefont {Park},\ and\ \citenamefont {Sim}}]{Lee2015}%
  \BibitemOpen
  \bibfield  {author} {\bibinfo {author} {\bibfnamefont {S.-S.~B.}\
  \bibnamefont {Lee}}, \bibinfo {author} {\bibfnamefont {J.}~\bibnamefont
  {Park}},\ and\ \bibinfo {author} {\bibfnamefont {H.-S.}\ \bibnamefont
  {Sim}},\ }\bibfield  {title} {\bibinfo {title} {{Macroscopic Quantum
  Entanglement of a Kondo Cloud at Finite Temperature}},\ }\href
  {https://doi.org/10.1103/PhysRevLett.114.057203} {\bibfield  {journal}
  {\bibinfo  {journal} {Phys. Rev. Lett.}\ }\textbf {\bibinfo {volume} {114}},\
  \bibinfo {pages} {057203} (\bibinfo {year} {2015})}\BibitemShut {NoStop}%
\bibitem [{\citenamefont {Alkurtass}\ \emph {et~al.}(2016)\citenamefont
  {Alkurtass}, \citenamefont {Bayat}, \citenamefont {Affleck}, \citenamefont
  {Bose}, \citenamefont {Johannesson}, \citenamefont {Sodano}, \citenamefont
  {S\o{}rensen},\ and\ \citenamefont {Le~Hur}}]{Alkurtass2016}%
  \BibitemOpen
  \bibfield  {author} {\bibinfo {author} {\bibfnamefont {B.}~\bibnamefont
  {Alkurtass}}, \bibinfo {author} {\bibfnamefont {A.}~\bibnamefont {Bayat}},
  \bibinfo {author} {\bibfnamefont {I.}~\bibnamefont {Affleck}}, \bibinfo
  {author} {\bibfnamefont {S.}~\bibnamefont {Bose}}, \bibinfo {author}
  {\bibfnamefont {H.}~\bibnamefont {Johannesson}}, \bibinfo {author}
  {\bibfnamefont {P.}~\bibnamefont {Sodano}}, \bibinfo {author} {\bibfnamefont
  {E.~S.}\ \bibnamefont {S\o{}rensen}},\ and\ \bibinfo {author} {\bibfnamefont
  {K.}~\bibnamefont {Le~Hur}},\ }\bibfield  {title} {\bibinfo {title}
  {{Entanglement structure of the two-channel Kondo model}},\ }\href
  {https://doi.org/10.1103/PhysRevB.93.081106} {\bibfield  {journal} {\bibinfo
  {journal} {Phys. Rev. B}\ }\textbf {\bibinfo {volume} {93}},\ \bibinfo
  {pages} {081106} (\bibinfo {year} {2016})}\BibitemShut {NoStop}%
\bibitem [{\citenamefont {Pari}\ \emph {et~al.}(2020)\citenamefont {Pari},
  \citenamefont {Garc\'{\i}a},\ and\ \citenamefont {Cornaglia}}]{Alvarez2020}%
  \BibitemOpen
  \bibfield  {author} {\bibinfo {author} {\bibfnamefont {N.~A.~A.}\
  \bibnamefont {Pari}}, \bibinfo {author} {\bibfnamefont {D.~J.}\ \bibnamefont
  {Garc\'{\i}a}},\ and\ \bibinfo {author} {\bibfnamefont {P.~S.}\ \bibnamefont
  {Cornaglia}},\ }\bibfield  {title} {\bibinfo {title} {{Quasiparticle Mass
  Enhancement as a Measure of Entanglement in the Kondo Problem}},\ }\href
  {https://doi.org/10.1103/PhysRevLett.125.217601} {\bibfield  {journal}
  {\bibinfo  {journal} {Phys. Rev. Lett.}\ }\textbf {\bibinfo {volume} {125}},\
  \bibinfo {pages} {217601} (\bibinfo {year} {2020})}\BibitemShut {NoStop}%
\bibitem [{\citenamefont {Kim}\ \emph {et~al.}(2021)\citenamefont {Kim},
  \citenamefont {Shim},\ and\ \citenamefont {Sim}}]{Kim2021}%
  \BibitemOpen
  \bibfield  {author} {\bibinfo {author} {\bibfnamefont {D.}~\bibnamefont
  {Kim}}, \bibinfo {author} {\bibfnamefont {J.}~\bibnamefont {Shim}},\ and\
  \bibinfo {author} {\bibfnamefont {H.-S.}\ \bibnamefont {Sim}},\ }\bibfield
  {title} {\bibinfo {title} {{Universal Thermal Entanglement of Multichannel
  Kondo Effects}},\ }\href {https://doi.org/10.1103/PhysRevLett.127.226801}
  {\bibfield  {journal} {\bibinfo  {journal} {Phys. Rev. Lett.}\ }\textbf
  {\bibinfo {volume} {127}},\ \bibinfo {pages} {226801} (\bibinfo {year}
  {2021})},\ \bibinfo {note} {https://arxiv.org/pdf/2112.01678}\BibitemShut
  {NoStop}%
\bibitem [{\citenamefont {Shim}\ \emph {et~al.}(2023)\citenamefont {Shim},
  \citenamefont {Kim},\ and\ \citenamefont {Sim}}]{Shim2023}%
  \BibitemOpen
  \bibfield  {author} {\bibinfo {author} {\bibfnamefont {J.}~\bibnamefont
  {Shim}}, \bibinfo {author} {\bibfnamefont {D.}~\bibnamefont {Kim}},\ and\
  \bibinfo {author} {\bibfnamefont {H.-S.}\ \bibnamefont {Sim}},\ }\bibfield
  {title} {\bibinfo {title} {{Hierarchical entanglement shells of multichannel
  Kondo clouds}},\ }\href {https://doi.org/10.1038/s41467-023-39234-6}
  {\bibfield  {journal} {\bibinfo  {journal} {Nature Communications}\ }\textbf
  {\bibinfo {volume} {14}},\ \bibinfo {pages} {3521} (\bibinfo {year}
  {2023})}\BibitemShut {NoStop}%
\bibitem [{\citenamefont {Nishikawa}\ and\ \citenamefont
  {Yoshioka}(2025)}]{Nishikawa2025}%
  \BibitemOpen
  \bibfield  {author} {\bibinfo {author} {\bibfnamefont {Y.}~\bibnamefont
  {Nishikawa}}\ and\ \bibinfo {author} {\bibfnamefont {T.}~\bibnamefont
  {Yoshioka}},\ }\bibfield  {title} {\bibinfo {title} {Quantum entanglement in
  a pure state of strongly correlated quantum impurity systems},\ }\href
  {https://doi.org/10.1103/PhysRevB.111.035112} {\bibfield  {journal} {\bibinfo
   {journal} {Phys. Rev. B}\ }\textbf {\bibinfo {volume} {111}},\ \bibinfo
  {pages} {035112} (\bibinfo {year} {2025})},\ \bibinfo {note}
  {http://arxiv.org/pdf/2404.18387}\BibitemShut {NoStop}%
\bibitem [{\citenamefont {Pustilnik}\ and\ \citenamefont
  {Glazman}(2001)}]{Pustilnik2001}%
  \BibitemOpen
  \bibfield  {author} {\bibinfo {author} {\bibfnamefont {M.}~\bibnamefont
  {Pustilnik}}\ and\ \bibinfo {author} {\bibfnamefont {L.~I.}\ \bibnamefont
  {Glazman}},\ }\bibfield  {title} {\bibinfo {title} {Kondo effect in real
  quantum dots},\ }\href {https://doi.org/10.1103/PhysRevLett.87.216601}
  {\bibfield  {journal} {\bibinfo  {journal} {Phys. Rev. Lett.}\ }\textbf
  {\bibinfo {volume} {87}},\ \bibinfo {pages} {216601} (\bibinfo {year}
  {2001})}\BibitemShut {NoStop}%
\bibitem [{\citenamefont {Potok}\ \emph {et~al.}(2007)\citenamefont {Potok},
  \citenamefont {Rau}, \citenamefont {Shtrikman}, \citenamefont {Oreg},\ and\
  \citenamefont {Goldhaber-Gordon}}]{Potok2007}%
  \BibitemOpen
  \bibfield  {author} {\bibinfo {author} {\bibfnamefont {R.~M.}\ \bibnamefont
  {Potok}}, \bibinfo {author} {\bibfnamefont {I.~G.}\ \bibnamefont {Rau}},
  \bibinfo {author} {\bibfnamefont {H.}~\bibnamefont {Shtrikman}}, \bibinfo
  {author} {\bibfnamefont {Y.}~\bibnamefont {Oreg}},\ and\ \bibinfo {author}
  {\bibfnamefont {D.}~\bibnamefont {Goldhaber-Gordon}},\ }\bibfield  {title}
  {\bibinfo {title} {{Observation of the two-channel Kondo effect}},\ }\href
  {https://doi.org/10.1038/nature05556} {\bibfield  {journal} {\bibinfo
  {journal} {Nature}\ }\textbf {\bibinfo {volume} {446}},\ \bibinfo {pages}
  {167} (\bibinfo {year} {2007})}\BibitemShut {NoStop}%
\bibitem [{\citenamefont {Parks}\ \emph {et~al.}(2010)\citenamefont {Parks},
  \citenamefont {Champagne}, \citenamefont {Costi}, \citenamefont {Shum},
  \citenamefont {Pasupathy}, \citenamefont {Neuscamman}, \citenamefont
  {Flores-Torres}, \citenamefont {Cornaglia}, \citenamefont {Aligia},
  \citenamefont {Balseiro}, \citenamefont {Chan}, \citenamefont {Abruña},\
  and\ \citenamefont {Ralph}}]{Parks2010}%
  \BibitemOpen
  \bibfield  {author} {\bibinfo {author} {\bibfnamefont {J.~J.}\ \bibnamefont
  {Parks}}, \bibinfo {author} {\bibfnamefont {A.~R.}\ \bibnamefont
  {Champagne}}, \bibinfo {author} {\bibfnamefont {T.~A.}\ \bibnamefont
  {Costi}}, \bibinfo {author} {\bibfnamefont {W.~W.}\ \bibnamefont {Shum}},
  \bibinfo {author} {\bibfnamefont {A.~N.}\ \bibnamefont {Pasupathy}}, \bibinfo
  {author} {\bibfnamefont {E.}~\bibnamefont {Neuscamman}}, \bibinfo {author}
  {\bibfnamefont {S.}~\bibnamefont {Flores-Torres}}, \bibinfo {author}
  {\bibfnamefont {P.~S.}\ \bibnamefont {Cornaglia}}, \bibinfo {author}
  {\bibfnamefont {A.~A.}\ \bibnamefont {Aligia}}, \bibinfo {author}
  {\bibfnamefont {C.~A.}\ \bibnamefont {Balseiro}}, \bibinfo {author}
  {\bibfnamefont {G.~K.-L.}\ \bibnamefont {Chan}}, \bibinfo {author}
  {\bibfnamefont {H.~D.}\ \bibnamefont {Abruña}},\ and\ \bibinfo {author}
  {\bibfnamefont {D.~C.}\ \bibnamefont {Ralph}},\ }\bibfield  {title} {\bibinfo
  {title} {{Mechanical Control of Spin States in Spin-1 Molecules and the
  Underscreened Kondo Effect}},\ }\href
  {https://doi.org/10.1126/science.1186874} {\bibfield  {journal} {\bibinfo
  {journal} {Science}\ }\textbf {\bibinfo {volume} {328}},\ \bibinfo {pages}
  {1370} (\bibinfo {year} {2010})},\ \Eprint
  {https://arxiv.org/abs/https://www.science.org/doi/pdf/10.1126/science.1186874}
  {https://www.science.org/doi/pdf/10.1126/science.1186874} \BibitemShut
  {NoStop}%
\bibitem [{\citenamefont {Minamitani}\ \emph {et~al.}(2012)\citenamefont
  {Minamitani}, \citenamefont {Tsukahara}, \citenamefont {Matsunaka},
  \citenamefont {Kim}, \citenamefont {Takagi},\ and\ \citenamefont
  {Kawai}}]{Minamitani2012}%
  \BibitemOpen
  \bibfield  {author} {\bibinfo {author} {\bibfnamefont {E.}~\bibnamefont
  {Minamitani}}, \bibinfo {author} {\bibfnamefont {N.}~\bibnamefont
  {Tsukahara}}, \bibinfo {author} {\bibfnamefont {D.}~\bibnamefont
  {Matsunaka}}, \bibinfo {author} {\bibfnamefont {Y.}~\bibnamefont {Kim}},
  \bibinfo {author} {\bibfnamefont {N.}~\bibnamefont {Takagi}},\ and\ \bibinfo
  {author} {\bibfnamefont {M.}~\bibnamefont {Kawai}},\ }\bibfield  {title}
  {\bibinfo {title} {{Symmetry-Driven Novel Kondo Effect in a Molecule}},\
  }\href {https://doi.org/10.1103/PhysRevLett.109.086602} {\bibfield  {journal}
  {\bibinfo  {journal} {Phys. Rev. Lett.}\ }\textbf {\bibinfo {volume} {109}},\
  \bibinfo {pages} {086602} (\bibinfo {year} {2012})}\BibitemShut {NoStop}%
\bibitem [{\citenamefont {Evers}\ \emph {et~al.}(2020)\citenamefont {Evers},
  \citenamefont {Koryt\'ar}, \citenamefont {Tewari},\ and\ \citenamefont {van
  Ruitenbeek}}]{Evers2020}%
  \BibitemOpen
  \bibfield  {author} {\bibinfo {author} {\bibfnamefont {F.}~\bibnamefont
  {Evers}}, \bibinfo {author} {\bibfnamefont {R.}~\bibnamefont {Koryt\'ar}},
  \bibinfo {author} {\bibfnamefont {S.}~\bibnamefont {Tewari}},\ and\ \bibinfo
  {author} {\bibfnamefont {J.~M.}\ \bibnamefont {van Ruitenbeek}},\ }\bibfield
  {title} {\bibinfo {title} {Advances and challenges in single-molecule
  electron transport},\ }\href {https://doi.org/10.1103/RevModPhys.92.035001}
  {\bibfield  {journal} {\bibinfo  {journal} {Rev. Mod. Phys.}\ }\textbf
  {\bibinfo {volume} {92}},\ \bibinfo {pages} {035001} (\bibinfo {year}
  {2020})}\BibitemShut {NoStop}%
\bibitem [{\citenamefont {Blesio}\ \emph {et~al.}(2024)\citenamefont {Blesio},
  \citenamefont {Manuel},\ and\ \citenamefont {Aligia}}]{Blesio2024}%
  \BibitemOpen
  \bibfield  {author} {\bibinfo {author} {\bibfnamefont {G.~G.}\ \bibnamefont
  {Blesio}}, \bibinfo {author} {\bibfnamefont {L.~O.}\ \bibnamefont {Manuel}},\
  and\ \bibinfo {author} {\bibfnamefont {A.~A.}\ \bibnamefont {Aligia}},\
  }\bibfield  {title} {\bibinfo {title} {Anisotropy-driven topological quantum
  phase transition in magnetic impurities},\ }\href
  {https://doi.org/10.1088/2633-4356/ada081} {\bibfield  {journal} {\bibinfo
  {journal} {Materials for Quantum Technology}\ }\textbf {\bibinfo {volume}
  {4}},\ \bibinfo {pages} {042001} (\bibinfo {year} {2024})}\BibitemShut
  {NoStop}%
\bibitem [{\citenamefont {Nozi{\`e}res}\ and\ \citenamefont
  {Blandin}(1980)}]{Nozieres1980}%
  \BibitemOpen
  \bibfield  {author} {\bibinfo {author} {\bibfnamefont {P.}~\bibnamefont
  {Nozi{\`e}res}}\ and\ \bibinfo {author} {\bibfnamefont {A.}~\bibnamefont
  {Blandin}},\ }\bibfield  {title} {\bibinfo {title} {{Kondo effect in real
  metals}},\ }\href {https://doi.org/10.1051/jphys:01980004103019300}
  {\bibfield  {journal} {\bibinfo  {journal} {{Journal de Physique}}\ }\textbf
  {\bibinfo {volume} {41}},\ \bibinfo {pages} {193} (\bibinfo {year}
  {1980})}\BibitemShut {NoStop}%
\bibitem [{\citenamefont {Amico}\ \emph {et~al.}(2008)\citenamefont {Amico},
  \citenamefont {Fazio}, \citenamefont {Osterloh},\ and\ \citenamefont
  {Vedral}}]{Amico2008}%
  \BibitemOpen
  \bibfield  {author} {\bibinfo {author} {\bibfnamefont {L.}~\bibnamefont
  {Amico}}, \bibinfo {author} {\bibfnamefont {R.}~\bibnamefont {Fazio}},
  \bibinfo {author} {\bibfnamefont {A.}~\bibnamefont {Osterloh}},\ and\
  \bibinfo {author} {\bibfnamefont {V.}~\bibnamefont {Vedral}},\ }\bibfield
  {title} {\bibinfo {title} {Entanglement in many-body systems},\ }\href
  {https://doi.org/10.1103/RevModPhys.80.517} {\bibfield  {journal} {\bibinfo
  {journal} {Rev. Mod. Phys.}\ }\textbf {\bibinfo {volume} {80}},\ \bibinfo
  {pages} {517} (\bibinfo {year} {2008})}\BibitemShut {NoStop}%
\bibitem [{\citenamefont {Laflorencie}(2016)}]{Laflorencie2016}%
  \BibitemOpen
  \bibfield  {author} {\bibinfo {author} {\bibfnamefont {N.}~\bibnamefont
  {Laflorencie}},\ }\bibfield  {title} {\bibinfo {title} {Quantum entanglement
  in condensed matter systems},\ }\href
  {https://doi.org/10.1016/j.physrep.2016.06.008} {\bibfield  {journal}
  {\bibinfo  {journal} {Physics Reports}\ }\textbf {\bibinfo {volume} {646}},\
  \bibinfo {pages} {1} (\bibinfo {year} {2016})},\ \bibinfo {note}
  {http://arxiv.org/pdf/1512.03388}\BibitemShut {NoStop}%
\bibitem [{\citenamefont {Wu}\ \emph {et~al.}(2004)\citenamefont {Wu},
  \citenamefont {Sarandy},\ and\ \citenamefont {Lidar}}]{WU-2004}%
  \BibitemOpen
  \bibfield  {author} {\bibinfo {author} {\bibfnamefont {L.-A.}\ \bibnamefont
  {Wu}}, \bibinfo {author} {\bibfnamefont {M.~S.}\ \bibnamefont {Sarandy}},\
  and\ \bibinfo {author} {\bibfnamefont {D.~A.}\ \bibnamefont {Lidar}},\
  }\bibfield  {title} {\bibinfo {title} {Quantum phase transitions and
  bipartite entanglement},\ }\href
  {https://doi.org/10.1103/physrevlett.93.250404} {\bibfield  {journal}
  {\bibinfo  {journal} {Physical Review Letters}\ }\textbf {\bibinfo {volume}
  {93}},\ \bibinfo {pages} {250404} (\bibinfo {year} {2004})}\BibitemShut
  {NoStop}%
\bibitem [{\citenamefont {Wagner}\ \emph {et~al.}(2018)\citenamefont {Wagner},
  \citenamefont {Chowdhury}, \citenamefont {Pixley},\ and\ \citenamefont
  {Ingersent}}]{Wagner2018}%
  \BibitemOpen
  \bibfield  {author} {\bibinfo {author} {\bibfnamefont {C.}~\bibnamefont
  {Wagner}}, \bibinfo {author} {\bibfnamefont {T.}~\bibnamefont {Chowdhury}},
  \bibinfo {author} {\bibfnamefont {J.~H.}\ \bibnamefont {Pixley}},\ and\
  \bibinfo {author} {\bibfnamefont {K.}~\bibnamefont {Ingersent}},\ }\bibfield
  {title} {\bibinfo {title} {Long-range entanglement near a kondo-destruction
  quantum critical point},\ }\href
  {https://doi.org/10.1103/PhysRevLett.121.147602} {\bibfield  {journal}
  {\bibinfo  {journal} {Phys. Rev. Lett.}\ }\textbf {\bibinfo {volume} {121}},\
  \bibinfo {pages} {147602} (\bibinfo {year} {2018})}\BibitemShut {NoStop}%
\bibitem [{\citenamefont {Blesio}\ \emph {et~al.}(2018)\citenamefont {Blesio},
  \citenamefont {Manuel}, \citenamefont {Roura-Bas},\ and\ \citenamefont
  {Aligia}}]{Blesio2018}%
  \BibitemOpen
  \bibfield  {author} {\bibinfo {author} {\bibfnamefont {G.~G.}\ \bibnamefont
  {Blesio}}, \bibinfo {author} {\bibfnamefont {L.~O.}\ \bibnamefont {Manuel}},
  \bibinfo {author} {\bibfnamefont {P.}~\bibnamefont {Roura-Bas}},\ and\
  \bibinfo {author} {\bibfnamefont {A.~A.}\ \bibnamefont {Aligia}},\ }\bibfield
   {title} {\bibinfo {title} {Topological quantum phase transition between
  {F}ermi liquid phases in an {A}nderson impurity model},\ }\href
  {https://doi.org/10.1103/physrevb.98.195435} {\bibfield  {journal} {\bibinfo
  {journal} {Physical Review B}\ }\textbf {\bibinfo {volume} {98}},\ \bibinfo
  {pages} {075434} (\bibinfo {year} {2018})}\BibitemShut {NoStop}%
\bibitem [{\citenamefont {Blesio}\ \emph {et~al.}(2019)\citenamefont {Blesio},
  \citenamefont {Manuel}, \citenamefont {Aligia},\ and\ \citenamefont
  {Roura-Bas}}]{Blesio2019}%
  \BibitemOpen
  \bibfield  {author} {\bibinfo {author} {\bibfnamefont {G.~G.}\ \bibnamefont
  {Blesio}}, \bibinfo {author} {\bibfnamefont {L.~O.}\ \bibnamefont {Manuel}},
  \bibinfo {author} {\bibfnamefont {A.~A.}\ \bibnamefont {Aligia}},\ and\
  \bibinfo {author} {\bibfnamefont {P.}~\bibnamefont {Roura-Bas}},\ }\bibfield
  {title} {\bibinfo {title} {Fully compensated {K}ondo effect for a two-channel
  spin {S}=1 impurity},\ }\href {https://doi.org/10.1103/physrevb.100.075434}
  {\bibfield  {journal} {\bibinfo  {journal} {Physical Review B}\ }\textbf
  {\bibinfo {volume} {100}},\ \bibinfo {pages} {075434} (\bibinfo {year}
  {2019})}\BibitemShut {NoStop}%
\bibitem [{\citenamefont {Žitko}\ \emph {et~al.}(2021)\citenamefont {Žitko},
  \citenamefont {Blesio}, \citenamefont {Manuel},\ and\ \citenamefont
  {Aligia}}]{Zitko2021}%
  \BibitemOpen
  \bibfield  {author} {\bibinfo {author} {\bibfnamefont {R.}~\bibnamefont
  {Žitko}}, \bibinfo {author} {\bibfnamefont {G.~G.}\ \bibnamefont {Blesio}},
  \bibinfo {author} {\bibfnamefont {L.~O.}\ \bibnamefont {Manuel}},\ and\
  \bibinfo {author} {\bibfnamefont {A.~A.}\ \bibnamefont {Aligia}},\ }\bibfield
   {title} {\bibinfo {title} {Iron phthalocyanine on {A}u(111) is a
  “non-{L}andau” {F}ermi liquid},\ }\href
  {https://doi.org/10.1038/s41467-021-26339-z} {\bibfield  {journal} {\bibinfo
  {journal} {Nature Communications}\ }\textbf {\bibinfo {volume} {12}},\
  \bibinfo {pages} {6027} (\bibinfo {year} {2021})}\BibitemShut {NoStop}%
\bibitem [{\citenamefont {Vojta}(2006)}]{Vojta2006}%
  \BibitemOpen
  \bibfield  {author} {\bibinfo {author} {\bibfnamefont {M.}~\bibnamefont
  {Vojta}},\ }\bibfield  {title} {\bibinfo {title} {Impurity quantum phase
  transitions},\ }\href {https://doi.org/10.1080/14786430500070396} {\bibfield
  {journal} {\bibinfo  {journal} {Philosophical Magazine}\ }\textbf {\bibinfo
  {volume} {86}},\ \bibinfo {pages} {1807} (\bibinfo {year}
  {2006})}\BibitemShut {NoStop}%
\bibitem [{\citenamefont {Fritz}\ and\ \citenamefont
  {Vojta}(2013)}]{Fritz2013}%
  \BibitemOpen
  \bibfield  {author} {\bibinfo {author} {\bibfnamefont {L.}~\bibnamefont
  {Fritz}}\ and\ \bibinfo {author} {\bibfnamefont {M.}~\bibnamefont {Vojta}},\
  }\bibfield  {title} {\bibinfo {title} {{The physics of Kondo impurities in
  graphene}},\ }\href {https://doi.org/10.1088/0034-4885/76/3/032501}
  {\bibfield  {journal} {\bibinfo  {journal} {Reports on Progress in Physics}\
  }\textbf {\bibinfo {volume} {76}},\ \bibinfo {pages} {032501} (\bibinfo
  {year} {2013})}\BibitemShut {NoStop}%
\bibitem [{\citenamefont {Moca}\ \emph {et~al.}(2021)\citenamefont {Moca},
  \citenamefont {Weymann}, \citenamefont {Werner},\ and\ \citenamefont
  {Zar\'and}}]{Moca2021}%
  \BibitemOpen
  \bibfield  {author} {\bibinfo {author} {\bibfnamefont {C.~P.}\ \bibnamefont
  {Moca}}, \bibinfo {author} {\bibfnamefont {I.}~\bibnamefont {Weymann}},
  \bibinfo {author} {\bibfnamefont {M.~A.}\ \bibnamefont {Werner}},\ and\
  \bibinfo {author} {\bibfnamefont {G.}~\bibnamefont {Zar\'and}},\ }\bibfield
  {title} {\bibinfo {title} {{Kondo Cloud in a Superconductor}},\ }\href
  {https://doi.org/10.1103/PhysRevLett.127.186804} {\bibfield  {journal}
  {\bibinfo  {journal} {Phys. Rev. Lett.}\ }\textbf {\bibinfo {volume} {127}},\
  \bibinfo {pages} {186804} (\bibinfo {year} {2021})}\BibitemShut {NoStop}%
\bibitem [{\citenamefont {Fishman}\ \emph {et~al.}(2022)\citenamefont
  {Fishman}, \citenamefont {White},\ and\ \citenamefont
  {Stoudenmire}}]{ITENSOR-PAPER}%
  \BibitemOpen
  \bibfield  {author} {\bibinfo {author} {\bibfnamefont {M.}~\bibnamefont
  {Fishman}}, \bibinfo {author} {\bibfnamefont {S.~R.}\ \bibnamefont {White}},\
  and\ \bibinfo {author} {\bibfnamefont {E.~M.}\ \bibnamefont {Stoudenmire}},\
  }\bibfield  {title} {\bibinfo {title} {{The ITensor Software Library for
  Tensor Network Calculations}},\ }\href
  {https://doi.org/10.21468/SciPostPhysCodeb.4} {\bibfield  {journal} {\bibinfo
   {journal} {SciPost Phys. Codebases}\ ,\ \bibinfo {pages} {4}} (\bibinfo
  {year} {2022})}\BibitemShut {NoStop}%
\bibitem [{\citenamefont {B\"usser}\ \emph {et~al.}(2010)\citenamefont
  {B\"usser}, \citenamefont {Martins}, \citenamefont {Costa~Ribeiro},
  \citenamefont {Vernek}, \citenamefont {Anda},\ and\ \citenamefont
  {Dagotto}}]{BUSSER-2010}%
  \BibitemOpen
  \bibfield  {author} {\bibinfo {author} {\bibfnamefont {C.~A.}\ \bibnamefont
  {B\"usser}}, \bibinfo {author} {\bibfnamefont {G.~B.}\ \bibnamefont
  {Martins}}, \bibinfo {author} {\bibfnamefont {L.}~\bibnamefont
  {Costa~Ribeiro}}, \bibinfo {author} {\bibfnamefont {E.}~\bibnamefont
  {Vernek}}, \bibinfo {author} {\bibfnamefont {E.~V.}\ \bibnamefont {Anda}},\
  and\ \bibinfo {author} {\bibfnamefont {E.}~\bibnamefont {Dagotto}},\
  }\bibfield  {title} {\bibinfo {title} {Numerical analysis of the spatial
  range of the {K}ondo effect},\ }\href
  {https://doi.org/10.1103/PhysRevB.81.045111} {\bibfield  {journal} {\bibinfo
  {journal} {Phys. Rev. B}\ }\textbf {\bibinfo {volume} {81}},\ \bibinfo
  {pages} {045111} (\bibinfo {year} {2010})}\BibitemShut {NoStop}%
\bibitem [{\citenamefont {Ribeiro}\ \emph {et~al.}(2019)\citenamefont
  {Ribeiro}, \citenamefont {Martins}, \citenamefont {G\'omez-Silva},\ and\
  \citenamefont {Anda}}]{RIBEIRO-2019}%
  \BibitemOpen
  \bibfield  {author} {\bibinfo {author} {\bibfnamefont {L.~C.}\ \bibnamefont
  {Ribeiro}}, \bibinfo {author} {\bibfnamefont {G.~B.}\ \bibnamefont
  {Martins}}, \bibinfo {author} {\bibfnamefont {G.}~\bibnamefont
  {G\'omez-Silva}},\ and\ \bibinfo {author} {\bibfnamefont {E.~V.}\
  \bibnamefont {Anda}},\ }\bibfield  {title} {\bibinfo {title} {Numerical study
  of the {K}ondo cloud using finite-{U} slave bosons},\ }\href
  {https://doi.org/10.1103/PhysRevB.99.085139} {\bibfield  {journal} {\bibinfo
  {journal} {Phys. Rev. B}\ }\textbf {\bibinfo {volume} {99}},\ \bibinfo
  {pages} {085139} (\bibinfo {year} {2019})}\BibitemShut {NoStop}%
\bibitem [{\citenamefont {Holzner}\ \emph {et~al.}(2009)\citenamefont
  {Holzner}, \citenamefont {McCulloch}, \citenamefont {Schollwöck},
  \citenamefont {von Delft},\ and\ \citenamefont
  {Heidrich-Meisner}}]{HOLZNER-2009}%
  \BibitemOpen
  \bibfield  {author} {\bibinfo {author} {\bibfnamefont {A.}~\bibnamefont
  {Holzner}}, \bibinfo {author} {\bibfnamefont {I.~P.}\ \bibnamefont
  {McCulloch}}, \bibinfo {author} {\bibfnamefont {U.}~\bibnamefont
  {Schollwöck}}, \bibinfo {author} {\bibfnamefont {J.}~\bibnamefont {von
  Delft}},\ and\ \bibinfo {author} {\bibfnamefont {F.}~\bibnamefont
  {Heidrich-Meisner}},\ }\bibfield  {title} {\bibinfo {title} {Kondo screening
  cloud in the single-impurity anderson model: A density matrix renormalization
  group study},\ }\href {https://doi.org/10.1103/physrevb.80.205114} {\bibfield
   {journal} {\bibinfo  {journal} {Physical Review B}\ }\textbf {\bibinfo
  {volume} {80}},\ \bibinfo {pages} {205114} (\bibinfo {year}
  {2009})}\BibitemShut {NoStop}%
\bibitem [{\citenamefont {Yang}(2005)}]{YANG-2005}%
  \BibitemOpen
  \bibfield  {author} {\bibinfo {author} {\bibfnamefont {M.-F.}\ \bibnamefont
  {Yang}},\ }\bibfield  {title} {\bibinfo {title} {Reexamination of
  entanglement and the quantum phase transition},\ }\href
  {https://doi.org/10.1103/PhysRevA.71.030302} {\bibfield  {journal} {\bibinfo
  {journal} {Phys. Rev. A}\ }\textbf {\bibinfo {volume} {71}},\ \bibinfo
  {pages} {030302} (\bibinfo {year} {2005})}\BibitemShut {NoStop}%
\bibitem [{\citenamefont {Bulla}\ \emph {et~al.}(2008)\citenamefont {Bulla},
  \citenamefont {Costi},\ and\ \citenamefont {Pruschke}}]{Bulla2008}%
  \BibitemOpen
  \bibfield  {author} {\bibinfo {author} {\bibfnamefont {R.}~\bibnamefont
  {Bulla}}, \bibinfo {author} {\bibfnamefont {T.~A.}\ \bibnamefont {Costi}},\
  and\ \bibinfo {author} {\bibfnamefont {T.}~\bibnamefont {Pruschke}},\
  }\bibfield  {title} {\bibinfo {title} {Numerical renormalization group method
  for quantum impurity systems},\ }\href
  {https://doi.org/10.1103/RevModPhys.80.395} {\bibfield  {journal} {\bibinfo
  {journal} {Rev. Mod. Phys.}\ }\textbf {\bibinfo {volume} {80}},\ \bibinfo
  {pages} {395} (\bibinfo {year} {2008})}\BibitemShut {NoStop}%
\bibitem [{\citenamefont {\ifmmode~\check{Z}\else \v{Z}\fi{}itko}\ and\
  \citenamefont {Pruschke}(2009)}]{Zitko2009}%
  \BibitemOpen
  \bibfield  {author} {\bibinfo {author} {\bibfnamefont {R.}~\bibnamefont
  {\ifmmode~\check{Z}\else \v{Z}\fi{}itko}}\ and\ \bibinfo {author}
  {\bibfnamefont {T.}~\bibnamefont {Pruschke}},\ }\bibfield  {title} {\bibinfo
  {title} {Energy resolution and discretization artifacts in the numerical
  renormalization group},\ }\href {https://doi.org/10.1103/PhysRevB.79.085106}
  {\bibfield  {journal} {\bibinfo  {journal} {Phys. Rev. B}\ }\textbf {\bibinfo
  {volume} {79}},\ \bibinfo {pages} {085106} (\bibinfo {year}
  {2009})}\BibitemShut {NoStop}%
\bibitem [{nrg(2023)}]{nrglj}%
  \BibitemOpen
  \href@noop {} {\bibinfo {title} {{N}{R}{G} {L}jubljana,
  \url{https://github.com/rokzitko/nrgljubljana} and
  \url{http://nrgljubljana.ijs.si/}}} (\bibinfo {year} {2023})\BibitemShut
  {NoStop}%
\bibitem [{SM()}]{SM}%
  \BibitemOpen
  \href@noop {} {}\bibinfo {note} {See Supplemental Material at http:... for
  the derivation of $\Sigma(x)$, a detailed analysis of the toy model, and
  additional information on the NRG calculations, which includes
  Refs.~\cite{Horodecki2009,Wilson1975,Bulla2008,Zitko2009,Ramsak2006}.}\BibitemShut
  {Stop}%
\bibitem [{\citenamefont {Horodecki}\ \emph {et~al.}(2009)\citenamefont
  {Horodecki}, \citenamefont {Horodecki}, \citenamefont {Horodecki},\ and\
  \citenamefont {Horodecki}}]{Horodecki2009}%
  \BibitemOpen
  \bibfield  {author} {\bibinfo {author} {\bibfnamefont {R.}~\bibnamefont
  {Horodecki}}, \bibinfo {author} {\bibfnamefont {P.}~\bibnamefont
  {Horodecki}}, \bibinfo {author} {\bibfnamefont {M.}~\bibnamefont
  {Horodecki}},\ and\ \bibinfo {author} {\bibfnamefont {K.}~\bibnamefont
  {Horodecki}},\ }\bibfield  {title} {\bibinfo {title} {Quantum entanglement},\
  }\href {https://doi.org/10.1103/RevModPhys.81.865} {\bibfield  {journal}
  {\bibinfo  {journal} {Rev. Mod. Phys.}\ }\textbf {\bibinfo {volume} {81}},\
  \bibinfo {pages} {865} (\bibinfo {year} {2009})}\BibitemShut {NoStop}%
\bibitem [{\citenamefont {Wilson}(1975)}]{Wilson1975}%
  \BibitemOpen
  \bibfield  {author} {\bibinfo {author} {\bibfnamefont {K.~G.}\ \bibnamefont
  {Wilson}},\ }\bibfield  {title} {\bibinfo {title} {The renormalization group:
  Critical phenomena and the kondo problem},\ }\href
  {https://doi.org/10.1103/RevModPhys.47.773} {\bibfield  {journal} {\bibinfo
  {journal} {Rev. Mod. Phys.}\ }\textbf {\bibinfo {volume} {47}},\ \bibinfo
  {pages} {773} (\bibinfo {year} {1975})}\BibitemShut {NoStop}%
\bibitem [{\citenamefont {Ram\ifmmode~\check{s}\else \v{s}\fi{}ak}\ \emph
  {et~al.}(2006{\natexlab{a}})\citenamefont {Ram\ifmmode~\check{s}\else
  \v{s}\fi{}ak}, \citenamefont {Mravlje}, \citenamefont
  {\ifmmode~\check{Z}\else \v{Z}\fi{}itko},\ and\ \citenamefont
  {Bon\ifmmode~\check{c}\else \v{c}\fi{}a}}]{Ramsak2006}%
  \BibitemOpen
  \bibfield  {author} {\bibinfo {author} {\bibfnamefont {A.}~\bibnamefont
  {Ram\ifmmode~\check{s}\else \v{s}\fi{}ak}}, \bibinfo {author} {\bibfnamefont
  {J.}~\bibnamefont {Mravlje}}, \bibinfo {author} {\bibfnamefont
  {R.}~\bibnamefont {\ifmmode~\check{Z}\else \v{Z}\fi{}itko}},\ and\ \bibinfo
  {author} {\bibfnamefont {J.}~\bibnamefont {Bon\ifmmode~\check{c}\else
  \v{c}\fi{}a}},\ }\bibfield  {title} {\bibinfo {title} {{Spin qubits in double
  quantum dots: Entanglement versus the Kondo effect}},\ }\href
  {https://doi.org/10.1103/PhysRevB.74.241305} {\bibfield  {journal} {\bibinfo
  {journal} {Phys. Rev. B}\ }\textbf {\bibinfo {volume} {74}},\ \bibinfo
  {pages} {241305} (\bibinfo {year} {2006}{\natexlab{a}})}\BibitemShut
  {NoStop}%
\bibitem [{\citenamefont {Gubernatis}\ \emph {et~al.}(1987)\citenamefont
  {Gubernatis}, \citenamefont {Hirsch},\ and\ \citenamefont
  {Scalapino}}]{Gubernatis1987}%
  \BibitemOpen
  \bibfield  {author} {\bibinfo {author} {\bibfnamefont {J.~E.}\ \bibnamefont
  {Gubernatis}}, \bibinfo {author} {\bibfnamefont {J.~E.}\ \bibnamefont
  {Hirsch}},\ and\ \bibinfo {author} {\bibfnamefont {D.~J.}\ \bibnamefont
  {Scalapino}},\ }\bibfield  {title} {\bibinfo {title} {Spin and charge
  correlations around an anderson magnetic impurity},\ }\href
  {https://doi.org/10.1103/PhysRevB.35.8478} {\bibfield  {journal} {\bibinfo
  {journal} {Phys. Rev. B}\ }\textbf {\bibinfo {volume} {35}},\ \bibinfo
  {pages} {8478} (\bibinfo {year} {1987})}\BibitemShut {NoStop}%
\bibitem [{\citenamefont {Costamagna}\ \emph {et~al.}(2006)\citenamefont
  {Costamagna}, \citenamefont {Gazza}, \citenamefont {Torio},\ and\
  \citenamefont {Riera}}]{Costamagna2006}%
  \BibitemOpen
  \bibfield  {author} {\bibinfo {author} {\bibfnamefont {S.}~\bibnamefont
  {Costamagna}}, \bibinfo {author} {\bibfnamefont {C.~J.}\ \bibnamefont
  {Gazza}}, \bibinfo {author} {\bibfnamefont {M.~E.}\ \bibnamefont {Torio}},\
  and\ \bibinfo {author} {\bibfnamefont {J.~A.}\ \bibnamefont {Riera}},\
  }\bibfield  {title} {\bibinfo {title} {{Anderson impurity in the
  one-dimensional Hubbard model for finite-size systems}},\ }\href
  {https://doi.org/10.1103/PhysRevB.74.195103} {\bibfield  {journal} {\bibinfo
  {journal} {Phys. Rev. B}\ }\textbf {\bibinfo {volume} {74}},\ \bibinfo
  {pages} {195103} (\bibinfo {year} {2006})}\BibitemShut {NoStop}%
\bibitem [{\citenamefont {Ram\ifmmode~\check{s}\else \v{s}\fi{}ak}\ \emph
  {et~al.}(2006{\natexlab{b}})\citenamefont {Ram\ifmmode~\check{s}\else
  \v{s}\fi{}ak}, \citenamefont {Sega},\ and\ \citenamefont
  {Jefferson}}]{RAMSAK-06}%
  \BibitemOpen
  \bibfield  {author} {\bibinfo {author} {\bibfnamefont {A.}~\bibnamefont
  {Ram\ifmmode~\check{s}\else \v{s}\fi{}ak}}, \bibinfo {author} {\bibfnamefont
  {I.}~\bibnamefont {Sega}},\ and\ \bibinfo {author} {\bibfnamefont {J.~H.}\
  \bibnamefont {Jefferson}},\ }\bibfield  {title} {\bibinfo {title}
  {Entanglement of two delocalized electrons},\ }\href
  {https://doi.org/10.1103/PhysRevA.74.010304} {\bibfield  {journal} {\bibinfo
  {journal} {Phys. Rev. A}\ }\textbf {\bibinfo {volume} {74}},\ \bibinfo
  {pages} {010304} (\bibinfo {year} {2006}{\natexlab{b}})}\BibitemShut
  {NoStop}%
\bibitem [{\citenamefont {Li}\ \emph {et~al.}(2017)\citenamefont {Li},
  \citenamefont {Feng},\ and\ \citenamefont {Dai}}]{Li2017}%
  \BibitemOpen
  \bibfield  {author} {\bibinfo {author} {\bibfnamefont {Y.}~\bibnamefont
  {Li}}, \bibinfo {author} {\bibfnamefont {X.-Y.}\ \bibnamefont {Feng}},\ and\
  \bibinfo {author} {\bibfnamefont {J.}~\bibnamefont {Dai}},\ }\bibfield
  {title} {\bibinfo {title} {{Local Kondo entanglement and symmetry protected
  local criticality in the two-impurity Kondo problem}},\ }\href
  {https://doi.org/10.1088/1742-6596/807/3/032004} {\bibfield  {journal}
  {\bibinfo  {journal} {Journal of Physics: Conference Series}\ }\textbf
  {\bibinfo {volume} {807}},\ \bibinfo {pages} {032004} (\bibinfo {year}
  {2017})}\BibitemShut {NoStop}%
\end{thebibliography}

\begin{thebibliography}{6}%
\makeatletter
\providecommand \@ifxundefined [1]{%
 \@ifx{#1\undefined}
}%
\providecommand \@ifnum [1]{%
 \ifnum #1\expandafter \@firstoftwo
 \else \expandafter \@secondoftwo
 \fi
}%
\providecommand \@ifx [1]{%
 \ifx #1\expandafter \@firstoftwo
 \else \expandafter \@secondoftwo
 \fi
}%
\providecommand \natexlab [1]{#1}%
\providecommand \enquote  [1]{``#1''}%
\providecommand \bibnamefont  [1]{#1}%
\providecommand \bibfnamefont [1]{#1}%
\providecommand \citenamefont [1]{#1}%
\providecommand \href@noop [0]{\@secondoftwo}%
\providecommand \href [0]{\begingroup \@sanitize@url \@href}%
\providecommand \@href[1]{\@@startlink{#1}\@@href}%
\providecommand \@@href[1]{\endgroup#1\@@endlink}%
\providecommand \@sanitize@url [0]{\catcode `\\12\catcode `\$12\catcode `\&12\catcode `\#12\catcode `\^12\catcode `\_12\catcode `\%12\relax}%
\providecommand \@@startlink[1]{}%
\providecommand \@@endlink[0]{}%
\providecommand \url  [0]{\begingroup\@sanitize@url \@url }%
\providecommand \@url [1]{\endgroup\@href {#1}{\urlprefix }}%
\providecommand \urlprefix  [0]{URL }%
\providecommand \Eprint [0]{\href }%
\providecommand \doibase [0]{https://doi.org/}%
\providecommand \selectlanguage [0]{\@gobble}%
\providecommand \bibinfo  [0]{\@secondoftwo}%
\providecommand \bibfield  [0]{\@secondoftwo}%
\providecommand \translation [1]{[#1]}%
\providecommand \BibitemOpen [0]{}%
\providecommand \bibitemStop [0]{}%
\providecommand \bibitemNoStop [0]{.\EOS\space}%
\providecommand \EOS [0]{\spacefactor3000\relax}%
\providecommand \BibitemShut  [1]{\csname bibitem#1\endcsname}%
\let\auto@bib@innerbib\@empty
\bibitem[S1]{SMHorodecki2009}%
  \BibitemOpen
  \bibfield  {author} {\bibinfo {author} {\bibfnamefont {R.}~\bibnamefont {Horodecki}}, \bibinfo {author} {\bibfnamefont {P.}~\bibnamefont {Horodecki}}, \bibinfo {author} {\bibfnamefont {M.}~\bibnamefont {Horodecki}},\ and\ \bibinfo {author} {\bibfnamefont {K.}~\bibnamefont {Horodecki}},\ }\href {https://doi.org/10.1103/RevModPhys.81.865} {\bibfield  {journal} {\bibinfo  {journal} {Rev. Mod. Phys.}\ }\textbf {\bibinfo {volume} {81}},\ \bibinfo {pages} {865} (\bibinfo {year} {2009})}\BibitemShut {NoStop}%
\bibitem [S2]{SMWilson1975}%
  \BibitemOpen
  \bibfield  {author} {\bibinfo {author} {\bibfnamefont {K.~G.}\ \bibnamefont {Wilson}},\ }\href {https://doi.org/10.1103/RevModPhys.47.773} {\bibfield  {journal} {\bibinfo  {journal} {Rev. Mod. Phys.}\ }\textbf {\bibinfo {volume} {47}},\ \bibinfo {pages} {773} (\bibinfo {year} {1975})}\BibitemShut {NoStop}%
\bibitem [S3]{SMBulla2008}%
  \BibitemOpen
  \bibfield  {author} {\bibinfo {author} {\bibfnamefont {R.}~\bibnamefont {Bulla}}, \bibinfo {author} {\bibfnamefont {T.~A.}\ \bibnamefont {Costi}},\ and\ \bibinfo {author} {\bibfnamefont {T.}~\bibnamefont {Pruschke}},\ }\href {https://doi.org/10.1103/RevModPhys.80.395} {\bibfield  {journal} {\bibinfo  {journal} {Rev. Mod. Phys.}\ }\textbf {\bibinfo {volume} {80}},\ \bibinfo {pages} {395} (\bibinfo {year} {2008})}\BibitemShut {NoStop}%
\bibitem [S4]{SMZitko2009}%
  \BibitemOpen
  \bibfield  {author} {\bibinfo {author} {\bibfnamefont {R.}~\bibnamefont {\ifmmode~\check{Z}\else \v{Z}\fi{}itko}}\ and\ \bibinfo {author} {\bibfnamefont {T.}~\bibnamefont {Pruschke}},\ }\href {https://doi.org/10.1103/PhysRevB.79.085106} {\bibfield  {journal} {\bibinfo  {journal} {Phys. Rev. B}\ }\textbf {\bibinfo {volume} {79}},\ \bibinfo {pages} {085106} (\bibinfo {year} {2009})}\BibitemShut {NoStop}%
\bibitem [S5]{SMnrglj}%
  \BibitemOpen
  \href@noop {} {\bibinfo {title} {{N}{R}{G} {L}jubljana, \url{https://github.com/rokzitko/nrgljubljana} and \url{http://nrgljubljana.ijs.si/}}} (\bibinfo {year} {2023})\BibitemShut {NoStop}%
\bibitem [S6]{SMRamsak2006}%
  \BibitemOpen
  \bibfield  {author} {\bibinfo {author} {\bibfnamefont {A.}~\bibnamefont {Ram\ifmmode~\check{s}\else \v{s}\fi{}ak}}, \bibinfo {author} {\bibfnamefont {J.}~\bibnamefont {Mravlje}}, \bibinfo {author} {\bibfnamefont {R.}~\bibnamefont {\ifmmode~\check{Z}\else \v{Z}\fi{}itko}},\ and\ \bibinfo {author} {\bibfnamefont {J.}~\bibnamefont {Bon\ifmmode~\check{c}\else \v{c}\fi{}a}},\ }\href {https://doi.org/10.1103/PhysRevB.74.241305} {\bibfield  {journal} {\bibinfo  {journal} {Phys. Rev. B}\ }\textbf {\bibinfo {volume} {74}},\ \bibinfo {pages} {241305} (\bibinfo {year} {2006})}\BibitemShut {NoStop}%
\end{thebibliography}

\end{document}